\documentclass[]{article}

\usepackage[utf8]{inputenc}
\usepackage{hyperref}
\usepackage{amsmath}
\usepackage{amsfonts}
\usepackage{amssymb}
\usepackage{tikz}
\usepackage{pgfplots}
\usepackage{psfrag}
\usepackage{import}
\usepackage{multirow}
\usepackage[utf8]{inputenc}  
\usepackage[T1]{fontenc}
\usepackage[french,english]{babel}   
\usepackage{graphicx}
\usepackage{verbatim}
\usepackage{tikz}
\usepackage{titlesec, blindtext, color}
\usepackage{lscape}
\usepackage{accents}
\usepackage{enumerate}
  \usepackage{amsthm}
  
  \newtheorem{dft}{Definition}
  \newtheorem{prop}{Proposition}
  \newtheorem{lemma}{Lemma}
  \usepackage[normalem]{ulem}
  \usepackage{cite}
\usepackage{eurosym}

\input{alphabet.tex}
\input{abrege.tex}
\newsavebox{\fminibox}
\newlength{\fminilength}
\newenvironment{fminipage}[1][\linewidth]
  {\setlength{\fminilength}{#1}
   \begin{lrbox}{\fminibox}\begin{minipage}{\fminilength}}
  {\end{minipage}\end{lrbox}\noindent\fbox{\usebox{\fminibox}}}

  \def\+{^\dagger}

\def\nequiv{\not\kern-.05em\equiv}
\def\egal{\kern-.5em=\kern-.5em}        
\def\propt{\kern-.2em\propto\kern-.2em} 

\def\intdouble{\int\kern-0.3em\int}
\def\inttriple{\int\kern-0.3em\int\kern-0.3em\int}

\def\rond#1{\overset{\kern-0.33em~_\circ}{#1}}
\def\rondit[#1]#2{\overset{\kern#1~_\circ}{#2}}

\def\babs{\begin{abstract}}             \def\eabs{\end{abstract}}
\def\barr{\begin{array}}                \def\earr{\end{array}}
\def\bcc{\begin{center}}                \def\ecc{\end{center}}

\def\bdes{\begin{description}}          \def\edes{\end{description}}
\def\bdoc{\begin{document}}             \def\edoc{\end{document}}
\def\ben{\begin{enumerate}}             \def\een{\end{enumerate}}
\def\beqn{\begin{eqnarray}}             \def\eeqn{\end{eqnarray}}
\def\beqnl#1{\beqn\label{#1}}           \def\eeqnl#1{\label{#1}\eeqn}
\def\beqnx{\begin{eqnarray*}}           \def\eeqnx{\end{eqnarray*}}
\def\bseqn{\begin{subeqnarray}}         \def\eseqn{\end{subeqnarray}}
\def\beq#1\eeq{\begin{equation}#1\end{equation}}
\def\bal#1\eal{\begin{align}#1\end{align}}
\def\balx#1\ealx{\begin{align*}#1\end{align*}}
\def\beqx{$$}                           \def\eeqx{$$}
\def\bfig{\protect\begin{figure}}       \def\efig{\protect\end{figure}}
\def\bfigx{\protect\begin{figure*}}     \def\efigx{\protect\end{figure*}}
\def\bfigt{\protect\begin{figurette}}   \def\efigt{\protect\end{figurette}}
\def\bfl{\begin{flushleft}}             \def\efl{\end{flushleft}}
\def\bfr{\begin{flushright}}            \def\efr{\end{flushright}}
\def\bit{\begin{itemize}}               \def\eit{\end{itemize}}
\def\bmi{\begin{minipage}}              \def\emi{\end{minipage}}
\def\bfmi{\begin{fminipage}}            \def\efmi{\end{fminipage}}
\def\bpic{\begin{picture}}              \def\epic{\end{picture}}
\def\bqu{\begin{quote}}                 \def\equ{\end{quote}}
\def\bqun{\begin{quotation}}            \def\equn{\end{quotation}}
\def\bsl{\begin{slide}}                 \def\esl{\end{slide}}
\def\btabb{\begin{tabbing}}             \def\etabb{\end{tabbing}}
\def\btabl{\begin{table}}               \def\etabl{\end{table}}
\def\btablx{\begin{table*}}             \def\etablx{\end{table*}}
\def\btab{\begin{tabular}} 
\def\btabu{\begin{tabular}}             \def\etabu{\end{tabular}}
\def\btabx{\begin{tabular*}}            \def\etabx{\end{tabular*}}
\def\bbib{\begin{thebibliography}{999}} \def\ebib{\end{thebibliography}}
\def\bver{\begin{verbatim}}             \def\ever{\end{verbatim}}
\def\bca{\begin{cases}}                  \def\eca{\end{cases}}

\newtheorem{condition}{\textbf{Condition}}
\def\AAkl{\mathcal{A}_{A,k,l}}
\def\Aunkl{\mathcal{A}_{1,k,l}}
\def\Adeuxkl{\mathcal{A}_{2,k,l}}
\def\Atroiskl{\mathcal{A}_{3,k,l}}
\def\sumCAnA{\sum_{i=1}^{n_A} C_t^{A,i}}
\def\sumCBnB{\sum_{i=1}^{n_B} C_t^{B,i}}
\def\sumCAk{\sum_{i=1}^{k} C_t^{A,i}}
\def\sumCBl{\sum_{i=1}^{l} C_t^{B,i}}
\def\sumCAkmoins{\sum_{i=1}^{k-1} C_t^{A,i}}
\def\sumCBlmoins{\sum_{i=1}^{l-1} C_t^{B,i}}
\def\CAmax{\bar{C}_t^A}
\def\CBmax{\bar{C}_t^B}
\def\Cetmax{\bar{C}_t^*}

\newcommand{\ubar}[1]{\underaccent{\bar}{#1}}

\title{Structural price model for electricity coupled markets}
\author{C. Alasseur, O. Féron}

\begin{document}

\maketitle

\begin{abstract}

We propose a new structural model that can compute the electricity spot and forward prices in two coupled markets with limited interconnection and multiple fuels. We choose a structural approach in order to represent some key characteristics of electricity spot prices such as their link to fuel prices, consumption level and production fleet. With this model, explicit formulas are also available for forward prices and other derivatives. We give some illustrative results of the behaviour of spot and forward prices, and of the values of transmission rights.
\\
\newline

\noindent \textbf{key words}: energy markets, structural models, derivatives pricing, electricity forwards, interconnection
\end{abstract}

\section{Introduction}

European electricity markets are mainly organized country by country. But these national markets are also interconnected with their neighbors. The two main advantages of interconnected markets are the following.
\bit
\item \textsl{Decreasing physical risk}. The interconnection between countries allows the pooling and sharing of available production capacities. And, because the electricity consumptions and production outages in these countries are not perfectly correlated, the interconnection makes the system more robust when facing extreme events. According to the French TSO\footnote{\textit{Introduction to interconnections}, http://clients.rte-france.com/lang/an/clients\_traders
	\_fournisseurs/services\_clients/dispositif\_global.jsp} 
, this is the first mission of interconnection: "These interconnections are therefore first used to ensure the operating safety of the power transmission networks."

\item \textsl{Optimizing financial cost.} The interconnection, by sharing production units in different markets, allows the (multi-country) system to be more efficient and to decrease the global (marginal) cost of electricity production.
\eit

The current trend in European electricity markets is integration. In November 2010, the Central West Europe (CWE) zone launched market coupling in the spot markets. This zone consists of five countries: France, Germany, Netherlands, Belgium, and Luxembourg. In 2014 and early 2015, the Euro area extended market coupling to a much larger zone which represents 19 European countries and around 85\% of the European demand\footnote{\textit{Market Coupling, A Major Step Towards Market Integration}, http://www.epexspot.com/en/market-coupling}. 

Under market coupling, players bid on spot markets through implicit auctions and do not have to take into account cross-border capacities. This is the market exchange which notifies players and use cross-border capacities in order to minimize the price difference within the zone. Therefore, market coupling enables to optimize cross-border interconnection usage.

As a consequence of market coupling, in 2015, the CWE region's spot price was unique 19\% of time which means that no congestion occurred at interconnections. In this case, the spot price was determined as if the five countries were only one. In 2014, market prices were the same in France and Germany around 50\% of the time, 27 \% in 2015. In May 2015, the methodology to calculate cross-border capacities in the CWE switched from price coupling to flow-based coupling in order to better take into account the network's physical constraints.\\

Therefore, modelling interconnected markets is more and more necessary to capture the (new) stylized facts of spot prices. This modelling is essential to efficiently participate in explicit long-term auctions of interconnection capacities organized by Joint Allocation Office JAO\footnote{http://www.jao.eu/main}. These auctions occur for many European interconnections for yearly and monthly delivery. Another application  of the model is to take advantage of a neighbouring market with higher liquidity to better hedge the risk of a particular market in which the liquidity is limited (proxy-hedging). Indeed, liquidity is very different on European markets. For example, traded volumes on both organized and OTC markets are around five times higher in Germany compared to other European markets (source \cite{EuropReport}). Market liquidity is measured in this report as the churn rate (i.e. the ratio between the global traded volume and electricity consumption). This rate is at least four times higher in Germany compared to others. The recent diminution of trading volumes in Belgium and Denmark is partly explained by players who use the more liquid German market for hedging instead of their domestic market \cite{Carr}.  Thus, for these applications, the targeted models must be able to represent forward contracts as well as spot prices. \\

 
 In the last few years, electricity markets have evolved very rapidly in terms of renewable production. According to the European Commission, the share of renewable production has grown from 14 \% in 2004 to 27.5 \% in 2014. As an example, wind and solar represented 16\% of German production in 2014 and their capacity has multiplied by 3.4 since 2004. This increase deeply changes the characteristics of electricity prices: average level, volatility, peaks, and seasonality. Especially, positive spikes on the electricity spot market have decreased in frequency and negative spikes are now quite common on spot and intraday markets. But transformations of the electricity production mix will continue in the near future: several thermal plants will be decommissioned and utilities are planning additional renewable capacities. Thus, electricity price models based on a purely statistical approach are difficult to calibrate and are not efficient in reflecting the future characteristics of electricity prices. This is why structural approaches, directly inspired by \cite{barlow2002diffusion}, are natural candidates to model electricity prices. These models are well-known for representing the stylized facts of spot prices and their (structural) link with fundamental factors like electricity demand, production capacities, and fuel prices. The most recent models \cite{aid2012structural,carmona2012survey} provide analytical formulas for forward prices which make them adaptable to risk management purposes in the case of a unique market. For example, \cite{Coulon2013} use a spike regime that they apply to the Texas market; \cite{carmona2013} propose possible changes in the merit order. \\
 
  But to our knowledge, only a few works aim at representing coupled markets. In particular, \cite{kustermann} and \cite{Fuss}, propose a simple model to represent two interconnected markets where they model the offer curve as an exponentially increasing function of the electricity demand. But they only  consider one fuel to produce electricity. The authors produce an analytical formula to retrieve the forward prices. \cite{MahringerFussProkopczuk} apply the model proposed in \cite{Fuss} to the valuation of transmission rights. 
  
In this paper, we propose a new structural model to represent the spot prices in two interconnected markets which account for an offer curve related to several production technologies. Under the classical assumption of no arbitrage, the forward price at maturity $T$ is equal to the expectation, under some risk-neutral probability, of the spot price at the future date $T$. We show that these forward prices can be computed by an quasi-analytical formula as well as call options and transmission rights which make the model well-adapted to risk management purposes.

The paper is organized as follows. Section 2 provides a description of the model and the construction of the electricity spot prices. In section 3 we give the analytical formulas for the valuation of derivatives. In section 4, we show some illustrative results based on an example. And section 5 concludes.

\section{Model description}

We propose a multi-commodity electricity price model for two interconnected zones. In this model, the technologies used to produce electricity can be specific to each zone. In this way, the model is able to represent the mix of different technologies for the two interconnected countries and the cross effects the interconnection induces. This model then can represent the peculiarities in the spot prices of the two markets. \\
In the following we assume a filtered probability space $(\Omega,\Fc, \{\Fc_t \},\Pbb) $, where $\{\Fc_t\}_t $ is the filtration generated by all of Wiener processes in the model.

\subsection{Spot price construction}
\label{sec_spot_price}
The model is directly inspired by \cite{aid2012structural} in the sense that at any time, the spot price can be related to the production cost of the marginal technology that is weighted by a scarcity function. Further, the marginal technology at time $t$ is the one, among all technologies in operation at time $t$, which has the highest production cost and which could provide one additional megawatt hour (MWh) of consumption. The scarcity function positively weights the marginal cost with respect to the distance from the total production capacity (i.e., sum of all the available capacities of production units where above this level the zone encounters production shortages) against the demand required by the consumers at that time. An exponential function is chosen for the scarcity function. \\

For each market $*=A,B$, we use the following notations:
\bit

\item $n_*$ is the number of available production technologies in market $*$,
\item $C^{*,k}_t$ is the available capacity in MW at time $t$ for the technology $k=1,\dots,n_*$,
\item $\Cb^*_t = \left( C_t^{*,k}\right)_{k=1,\dots,n_*}$ is the set of available capacities at time $t$,
\item $\Cetmax=\sum_{l=1}^{n_*} C^{*,l}_t$ is the total available capacity,
\item $D^*_t$ is the electricity demand at time $t$,
\item $S^{*,k}_t$ is the production cost in \euro/MWh at time $t$ of the technology $k=1,\dots,n_*$,
\item $\Sb^*_t = \left( S_t^{*,k}\right)_{k=1,\dots,n_*}$ is the set of production costs at time $t$.
\eit

In this section we suppose, for notation simplicity, that the production costs are sorted at time $t$ for each market, that is $S^{*,1}_t \le S^{*,2}_t \le \dots \le S^{*,n_*}_t $, $*=A,B$. The case of possible switches in the merit order is addressed in section \ref{sec_switch}. We define the marginality intervals as:
\beqx
I_t^{*,k}  = \left[\sum_{i=0}^{k-1}C^{*,i}_t ~;~ \sum_{i=0}^{k}C^{*,i}_t\right], \quad *=A,B  
\eeqx
with the additional bound assumption of $C^{A,0}_t=C^{B,0}_t=0$.

We define the offer curve $P^*$ for each market $*=A,B$ as a piecewise exponential function of the power demand. Given a demand $d^*$ to satisfy in market $*$, the offer function is given by:
\beq
P^*(\Sb^*_t,\Cb^*_t,d^*)=\sum_{k=1}^{n_*}f^*(S^{*,k}_t,\bar{C}^*_t,d^*)\unbb_{d\in I^{*,k}_t}
\eeq
with
\beq
\label{eq_exp_func}
f^*(s,c,d)=s e^{\alpha^* + \beta^* 
	\left(c - d\right)}
\eeq
where the fixed coefficients are $\alpha^*\in \Rbb$ and $\beta^* \le 0$, $*=A,B$. The conditions that these coefficients are not dependent on  technology $k$ and that $\beta^*$ is negative, ensure that the offer curve is an increasing (but not necessarily continuous) function of the demand level. However these coefficients can be different in each market. In the case where the markets are disconnected, $P^*(\Sb^*_t,\Cb^*_t,d^*)$ corresponds to the spot price in the market $*$. 

Next, the commercial flow $E_t$ through the interconnection between zone $A$ and zone $B$.\footnote{$E_t>0$ means zone $A$ exports power to zone $B$ and $E_t<0$ means zone $B$ exports power to zone $A$.} Because of the coupling mechanism, the commercial flow $E_t$ is determined such as to minimize the absolute distance between the spot prices of the two countries. Furthermore, $E_t$ is bounded by the NTC (Net Transfer Capacity) $\left[\ubar{E}, \bar{E} \right]$ set by network operators (note that $\ubar{E} \leq 0$ and $\bar{E} \geq 0$). The NTC is determined by network operators driven by physical constraints: the physical capacity of the interconnection line and the additional operational network constraints to guarantee the sufficient reliability of the system.\\

Given the optimal commercial flow $E_t$, the demand to satisfy in zone $A$ is $D^A_t + E_t$ whereas the demand to satisfy in zone $B$ is $D_t^B - E_t$. The optimal commercial flow $E_t$ can be defined as follows:
\begin{dft}
	\label{def_Et}
	The optimal flow $E_t$ is defined as: 
\bit
\item if $P^A(\Sb_t^A,\Cb^A_t,D^A_t)\le P^B(\Sb_t^B,\Cb^B_t,D^B_t)$ then
\beqnx
E_t = &&\sup_{\left[ \ubar{E}~;~ \bar{E} \right]} e \\
& & s.t.\\
& & ~ P^A(\Sb_t^A,\Cb^A_t,D^A_t+e)\le P^B(\Sb_t^B,\Cb^B_t,D^B_t-e)
\eeqnx
\item if $P^A(\Sb_t^A,\Cb^A_t,D^A_t)\ge P^B(\Sb_t^B,\Cb^B_t,D^B_t)$ then
\beqnx
E_t = &&\inf_{\left[ \ubar{E}~;~ \bar{E} \right]} e \\
& & s.t.\\
& & ~ P^A(\Sb_t^A,\Cb^A_t,D^A_t+e)\ge P^B(\Sb_t^B,\Cb^B_t,D^B_t-e)
\eeqnx
\eit 
\end{dft}

The power spot prices depend on the optimal flow $E_t$ between the two countries. As suggested in \cite{kustermann}, we partition the state space into three events\footnote{for simplicity the dependence in time $t$ is implicit in all the events defined in this section} depending on the saturated and coupling situations of the interconnection 
%
%
%
:

\begin{itemize}
	\item saturated interconnection from zone $A$ to $B$, non-coupling situation: \\
	$\Ac_1 := \{w \in \Omega : P^A(\Sb_t^A,\Cb_t^A, D_t^A+\bar{E})\leq P^B(\Sb_t^B,\Cb_t^B, D_t^B-\bar{E})\}$ 
	
	\item saturated interconnection from zone $B$ to $A$, non-coupling situation: \\
	$\Ac_2 := \{w \in \Omega : P^A(\Sb_t^A,\Cb_t^A, D_t^A+\ubar{E})\geq P^B(\Sb_t^B,\Cb_t^B, D_t^B-\ubar{E})\}$
	\item non-saturated interconnection and coupled markets: \\
	$\Ac_3 := \Omega \backslash (\Ac_1\cup \Ac_2)$
\end{itemize}

Under the events $\Ac_1$ and $\Ac_2$, the prices in the two markets remain different even after using the interconnection at the maximum of its capacity. Under the event $\Ac_3$ we consider the prices in the two markets to be equal and this common price has to be determined. \\
%
In the proposed model, we consider several production technologies. Therefore we need to refine the previous partition by introducing the marginality in the two countries. The $\Mc_{k,l} $ is defined as the event when at time $t$, the technology $k$ is marginal in market $A$ and the technology $l$ is marginal in market $B$. In the case where the production costs are sorted this event can be defined as:
\beqx
\Mc_{k,l} := \{\omega \in \Omega: ~D_t^A + E_t \in I_t^{A,k}~; ~ D_t^B - E_t \in I_t^{B,l}\}
\eeqx 
In the case of possible switches, as we will see in section \ref{sec_switch}, this event is not easily defined but this does not create any difficulty in the computation results. \\

Moreover, we define $\Ac_{i,k,l}  = \Ac_i \cap \Mc_{k,l}$ so that the state space can be partition as follows:
\beqx
\Omega = \bigcup\limits_{1\le i \le 3,~1\le k \le,n_A,~1 \le l \le n_B} \Ac_{i,k,l}
\eeqx

%
If $\Vb_t=\left(\Sb_t^A,\Cb_t^A,D_t^A,\Sb_t^B,\Cb_t^B,D_t^B \right)$ is the whole set of variables at time $t$, then the power spot prices $\bar{P}^*\left(\Vb_t \right)$, $*=A,B$, can be determined using this partition:


\begin{equation}
\label{spotA}
\begin{split}
\bar{P}^A\left(\Vb_t \right)  = \sum_{k=1}^{n_A} \sum_{l=1}^{n_B} & f^A(S^{A,k}_t,\bar{C}^A_t,D^A_t+\bar{E}) \unbb_{\Ac_{1,k,l}} + \\
& f^A(S^{A,k}_t,\bar{C}^A_t,D^A_t+\ubar{E}) \unbb_{\Ac_{2,k,l}} +\\\\
& f^{A,B}(\Vb_t)\unbb_{\Ac_{3,k,l}}
\end{split}
\end{equation}

and 

\begin{equation}
\label{spotB}
\begin{split}
\bar{P}^B\left(\Vb_t \right)  = \sum_{k=1}^{n_A} \sum_{l=1}^{n_B} 
& f^B(S^{B,l}_t,\bar{C}^B_t,D^B_t-\bar{E}) \unbb_{\Ac_{1,k,l}} + \\
& f^B(S^{B,l}_t,\bar{C}^B_t,D^B_t-\ubar{E}) \unbb_{\Ac_{2,k,l}} +\\\\
& f^{A,B}(\Vb_t)\unbb_{\Ac_{3,k,l}}
\end{split}
\end{equation}

with $f^A$ and $f^B$ defined in equation \eqref{eq_exp_func}. And $f^{A,B}$ is the common price of the two markets under the event $\Ac_{3,k,l}$ that has to be determined.

\hspace{1cm}

 \subsection{Optimal $E_t$ and corresponding prices}
 In this section we precisely determine the terms of equations \eqref{spotA} and \eqref{spotB}. The cases $\Ac_{1,k,l}$ and $\Ac_{2,k,l}$ are quite trivial and very similar to the cases described in \cite{kustermann}. The event $\Ac_{3,k,l}$ is more tricky because of the discontinuity of the offer curves, both the optimal flow $E_t$ and the common price $f^{A,B}$ have to be determined. 
 \subsubsection{Cases $\Ac_{1,k,l}$ and $\Ac_{2,k,l}$}
 The events $\Ac_{1,k,l}$ and $\Ac_{2,k,l}$ correspond to the saturated situations where zone $A$ ($B$) exports to zone $B$ ($A$) at the maximum of the interconnection capacity. That is, $\bar{E}$ (resp. $\ubar{E}$), and the technology $k$ is marginal in market $A$ and technology $l$ is marginal in market $B$. By these event's definition, we have:
 \beq
 \Vb_t \in \Ac_{1,k,l} \quad \Leftrightarrow \quad \left\{ \begin{matrix} 
 D^A_t + \bar{E} \in I^{A,k}_t \\ \\
 D^B_t - \bar{E} \in I^{B,l}_t \\ \\
 f^A(S^{A,k}_t,\CAmax,D^A_t+\bar{E}) \le f^B(S^{B,l}_t,\CBmax,D^B_t-\bar{E})
 \end{matrix}
 \right. 
 \eeq
 
 and 
  \beq
  \Vb_t \in \Ac_{2,k,l} \quad \Leftrightarrow \quad \left\{ \begin{matrix} 
  	D^A_t + \ubar{E} \in I^{A,k}_t \\ \\
  	D^B_t - \ubar{E} \in I^{B,l}_t \\ \\
  	f^A(S^{A,k}_t,\CAmax,D^A_t+\ubar{E}) \ge f^B(S^{B,l}_t,\CBmax,D^B_t-\ubar{E})
  \end{matrix}
  \right. 
  \eeq
 
%
%
%
%
%
 \subsubsection{Case $\mathcal{A}_{3,k,l}$} 
 Under $\Ac_{3,k,l}$, the commercial flow is not saturated, that is, $\ubar{E}<E_t<\bar{E}$. This condition means that the prices in zones $A$ and $B$ have converged to price $f^{A,B}(\Vb_t)$.
 
 Because of the discontinuity of the offer curves, there are several cases to define the optimal commercial flow $E_t$. Indeed, in the case where $E_t$ reaches a discontinuity point of one of the offer curves, the absolute distance between $P^A(\Sb^A_t, \Cb^A_t,D^A_t+E_t)$ and $P^B(\Sb^B_t, \Cb^B_t,D^B_t-E_t)$ may be not zero. 
 
\begin{figure}[htb!]
 	\begin{center}
\begin{tabular}{cc}
\includegraphics[scale=0.4]{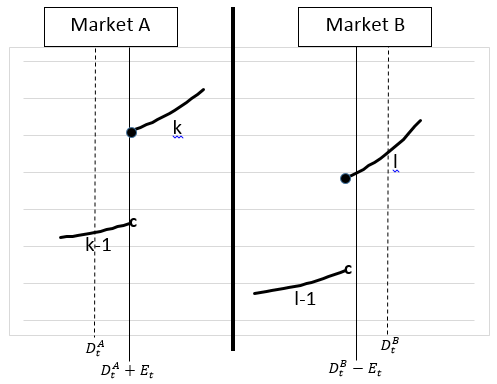} 	
&
\includegraphics[scale=0.4]{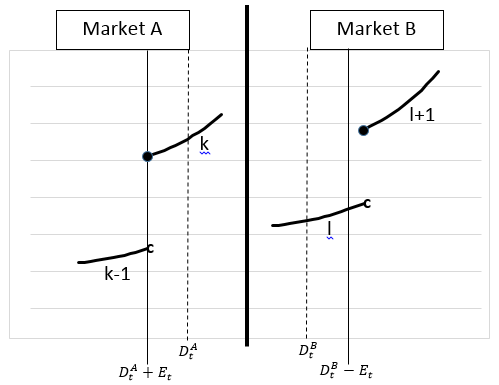} 	
\\
$\Ac_{3,k,l}^A$ with $E_t>0$ & $\Ac_{3,k,l}^A$ with $E_t<0$
\\ \\
\includegraphics[scale=0.4]{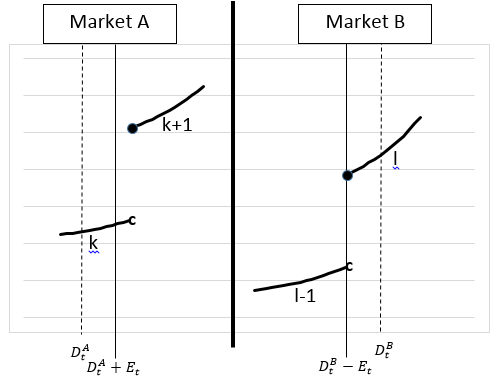} 	
&
\includegraphics[scale=0.4]{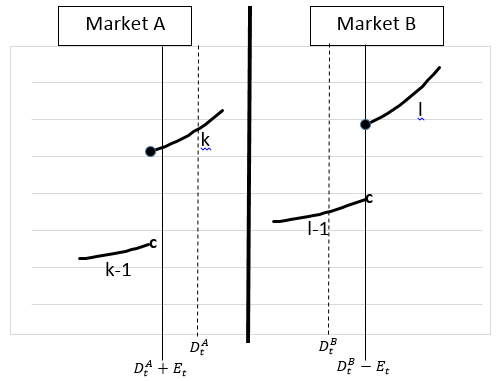} 	
\\
$\Ac_{3,k,l}^B$ with $E_t>0$ & $\Ac_{3,k,l}^B$ with $E_t<0$
	\end{tabular} 	
	\caption{\label{fig_scheme}Schematic representation of events $\Ac_{3,k,l}^A$ and $\Ac_{3,k,l}^B$ where $E_t$ reaches a point of discontinuity.
		}
		\end{center}
\end{figure}

  We decompose $\Ac_{3,k,l}$ into 3 incompatible subsets depending on the discontinuity points of the offer curves:
 \bit
  \item $\Ac_{3,k,l}^A=\{\omega \in \Ac_{3,k,l}: D^A_t+E_t$ is a point of discontinuity of  $P^A\}$
   \item $\Ac_{3,k,l}^B=\{\omega \in \Ac_{3,k,l}: D^B_t-E_t$ is a point of discontinuity of  $P^B\}$
  \item $\Ac_{3,k,l}^C=\Ac_{3,k,l} \backslash(\Ac_{3,k,l}^A \cup  \Ac_{3,k,l}^B ) $
 \eit
 These events then depend on the fact that $E_t$ reaches a discontinuity point on one of the offer curves. Figure \ref{fig_scheme} illustrates the events $\Ac_{3,k,l}^A$ and $\Ac_{3,k,l}^B$ which depend on the sign of $E_t$.
 
The Propositions \ref{prop_Ca} to \ref{prop_Cc} determine the events $\Ac_{3,k,l}^A$, $\Ac_{3,k,l}^B$ and $\Ac_{3,k,l}^C$, respectively, in terms of the inequalities in $\Vb_t$.
\begin{prop}
	\label{prop_Ca}
Set $g(\Vb_t)=\sum_{i=0}^{k-1}C^{A,i}_t - D^A_t$. 

	\noindent
The $\Vb_t \in \Ac_{3,k,l}^A$ is equivalent to the following inequalities:
\begin{enumerate}[{P\ref{prop_Ca}}-1 :]
\item $\ubar{E}< g(\Vb_t) < \bar{E}$ \label{prop1_H1}
\item $D^B_t - g(\Vb_t) \in I^{B,l}_t$ \label{prop1_H2}
\item $f^A \left(S^{A,k-1}_t,\bar{C}^A_t,D^A_t+g(\Vb_t) \right) \le f^B \left(S^{B,l}_t,\bar{C}^B_t,D^B_t-g(\Vb_t) \right)$ \label{prop1_H3}
\item $f^A \left(S^{A,k}_t,\bar{C}^A_t,D^A_t+g(\Vb_t) \right) \ge f^B \left(S^{B,l}_t,\bar{C}^B_t,D^B_t-g(\Vb_t) \right)$\label{prop1_H4}
\end{enumerate}
Moreover, if $\Vb_t \in \Ac_{3,k,l}^A$, then the common price is:
\beqx
f^{A,B}(\Vb_t)=f^B(S^{B,l}_t,\bar{C}^B_t,D^B_t-g(\Vb_t))
\eeqx

\end{prop}

%

\begin{prop}
	\label{prop_Cb}
	Set $g(\Vb_t)=D^B_t- \sum_{i=0}^{l-1}C^{B,i}_t$. 
	
	\noindent
	The $\Vb_t \in \Ac_{3,k,l}^B$ is equivalent to the following inequalities:
	\begin{enumerate}[{P\ref{prop_Cb}}-1 :]
		\item $\ubar{E} < g(\Vb_t) < \bar{E}$ \label{prop2_H1}
		\item $D^A_t + g(\Vb_t) \in I^{A,k}_t$ \label{prop2_H2}
		\item $f^B \left(S^{B,l-1}_t,\bar{C}^B_t,D^B_t-g(\Vb_t) \right) \le  f^A \left(S^{A,k}_t,\bar{C}^A_t,D^A_t+g(\Vb_t) \right) $ \label{prop2_H3}
		\item $f^B \left(S^{B,l}_t,\bar{C}^B_t,D^B_t-g(\Vb_t) \right) \ge f^A \left(S^{A,k}_t,\bar{C}^A_t,D^A_t+g(\Vb_t) \right) $ \label{prop2_H4}
	\end{enumerate}
	Moreover, if $\Vb_t \in \Ac_{3,k,l}^B$, then the common price is
	\beqx
	f^{A,B}(\Vb_t)=f^A(S^{A,k}_t,\bar{C}^A_t,D^A_t+g(\Vb_t))
	\eeqx
\end{prop}

%

\begin{prop}
	\label{prop_Cc}
	Set $g(\Vb_t)=\frac{\ln S_t^{A,k} -\ln S_t^{B,l} + \alpha^A - \alpha^B+  \beta^A(\bar{C}_t^{A}- D_t^A ) - \beta^B(\bar{C}_t^{B}- D_t^B)}{\beta^A + \beta^B}$. 
	
	The $\Vb_t \in \Ac_{3,k,l}^{C}$ is equivalent to the following inequalities:
	\begin{enumerate}[{P\ref{prop_Cc}}-1 :]
		\item $\ubar{E} < g(\Vb_t) < \bar{E}$
		\item $\sum_{i=0}^{k-1} C_t^{A,i} < D^A_t + g(\Vb_t) < \sum_{i=0}^{k} C_t^{A,i}$
		\item $\sum_{i=0}^{l-1} C_t^{B,i} <D^B_t - g(\Vb_t) <\sum_{i=0}^{l} C_t^{B,i}$
	\end{enumerate}
	Moreover, if $\Vb_t \in \Ac_{3,k,l}^C$, then the common price is:
	\beqx
	f^{A,B}(\Vb_t)=f^A(S^{A,k}_t,\bar{C}^A_t,D^A_t+g(\Vb_t))=f^B(S^{B,l}_t,\bar{C}^B_t,D^B_t-g(\Vb_t))
	\eeqx
\end{prop}
\noindent
The proof of Propositions \ref{prop_Ca}, \ref{prop_Cb} and \ref{prop_Cc} are given in appendix \ref{sec_proof}.



%

\subsection{The case of switching}
\label{sec_switch}
The previous section described the construction of the spot prices for fixed fuels. Now we consider possible switches in the fuels, that is permutations 
$\pi=(\pi^A,\pi^B)$ 
with 
$\pi^*=(\pi^*_{1},\pi^*_{2}\dots,\pi^{*}_{n_*})$
, where $*=A,B$ such that, at time $t$, we have 
$S^{*,\pi^*_{1}}_t<S^{*,\pi^*_{2}}_t<\dots <S^{*,\pi^*_{n_*}}_t$
. The merit order event is:
\beqx
\Sc^{\pi}=\left\{\omega \in \Omega ~;~S^{*,\pi^*_{1}}_t<S^{*,\pi^*_{2}}_t<\dots <S^{*,\pi^*_{n_*}}_t,~*=A,B\right\}
\eeqx


The marginality event $\Mc_{k,l}$ cannot be directly written in terms of only one interval. Indeed, $\Mc_{k,l}$ is the event where technology $k$ is marginal in zone $A$ and technology $l$ is marginal in zone $B$. The event then is a union that depends on the merit order:
\beq
\label{Mkl_switch}
\Mc_{k,l}= \bigcup_{\pi \in \Pi}\left( \Sc^\pi \cap \left\{\omega \in \Omega: D^A_t \in I_t^{A,\pi^A_k} ~;~ D^B_t \in I_t^{B,\pi^B_l} \right\} \right)
\eeq
with $\Pi$ as the whole set of possible permutations and:
\beqx
I_t^{*,\pi^*_k}  = \left[\sum_{i=0}^{\pi^*_{k-1}}C^{*,\pi^*_i}_t ~;~ \sum_{i=0}^{\pi^*_{k}}C^{*,\pi^*_i}_t\right], \quad *=A,B  
\eeqx
With the definition of $\Mc_{k,l}$ and $\Ac_{i,k,l}=\Ac_i \cap \Mc_{k,l}$, the spot prices in the two markets are then defined by:
\begin{equation}
\label{spotAswitch}
\begin{split}
\bar{P}^A\left(\Vb_t \right)  = \sum_{\pi \in \Pi}\sum_{k=0}^{n_A} \sum_{l=0}^{n_B} & f^A(S^{A,\pi_{k}^A}_t,\bar{C}^A_t,D^A_t+\bar{E}) \unbb_{\Ac_{1,k,l}}(\Vb_t) \unbb_{\Sc^{\pi}}(\Vb_t) + \\
& f^A(S^{A,\pi_{k}^A}_t,\bar{C}^A_t,D^A_t+\ubar{E}) \unbb_{\Ac_{2,k,l}}(\Vb_t) \unbb_{\Sc^{\pi}}(\Vb_t) +\\\\
& f^{A,B}(\Vb_t)\unbb_{\Ac_{3,k,l}}(\Vb_t) \unbb_{\Sc^{\pi}}(\Vb_t)
\end{split}
\end{equation}

and 

\begin{equation}
\label{spotBswitch}
\begin{split}
\bar{P}^B\left(\Vb_t \right)  = \sum_{\pi \in \Pi}\sum_{k=0}^{n_A} \sum_{l=0}^{n_B} 
& f^B(S^{B,\pi_{l}^B}_t,\bar{C}^B_t,D^B_t-\bar{E}) \unbb_{\Ac_{1,k,l}}(\Vb_t) \unbb_{\Sc^{\pi}}(\Vb_t) + \\
& f^B(S^{B,\pi_{l}^B}_t,\bar{C}^B_t,D^B_t-\ubar{E}) \unbb_{\Ac_{2,k,l}}(\Vb_t) \unbb_{\Sc^{\pi}}(\Vb_t) +\\\\
& f^{A,B}(\Vb_t)\unbb_{\Ac_{3,k,l}}(\Vb_t) \unbb_{\Sc^{\pi}}(\Vb_t)
\end{split}
\end{equation}
with $f^{A,B}(\Vb_t)$ having the same expression as in the previous section, which replaces $k$ and $l$ with $\pi_k^A$ and $\pi_l^B$, respectively.

\subsection{Models for fuel prices and demands in the Risk-Neutral probability}
\label{sec_model_fuel}
Our setting is clearly an incomplete market (indeed demand and capacities, if random, are non-tradable). Consequently, the market has an infinite number of Equivalent Martingale Measures. In the following we directly consider the model under a pricing probability $\Qbb$ such that, as in \cite{kustermann} or \cite{aid2012structural}, it coincides with the physical measure for the non-tradable risk factors (demands and production capacities). \\

For simplicity, we consider the deterministic production capacities defined by:

\begin{equation}
C_t^{*,k} = f_t^{*,k} \quad \forall k \in \{1,...,n_*\}, * = A,B 
\end{equation}
with \[f_t^{*,k} = \Cc_k^{*,k}\cos(2\pi t) + \Cc_2^{*,k}\sin(2\pi t)\]
However, nothing prevents us from considering the stochastic capacities with Wiener processes, only the complexity induced by adding new random variables.\\

The demand processes are modeled as in \cite{aid2012structural} with a deterministic component and an Ornstein-Ulhenbeck process:
\begin{equation}
D_t^* = f_t^* + \tilde{D}_t^*, \quad * = A,B
\end{equation}  
with 
\[f_t^* = \mathcal{X}_1^*\cos(2\pi t) + X_2^*\sin(2\pi t), \text{ } * = A,B \]
and \[ d\tilde{D}_t^* = -a^*\tilde{D}_t^* dt + \sigma^* dW_t^*\]

If $N$ is the total number of technologies. Then, we consider the possible case where one technology is available in both zones: $N\le n_A + n_B$. The production costs are modelled by a log-normal Ornstein-Uhlenbeck process:
\begin{equation}
d\log S_t^{n} = a_n(m_n(t) - \log S_t^{n})dt + \sigma_n dW_t^n,\quad  n = 1,\dots,N 
\end{equation}
All of these processes can be correlated by a $(N+2) \times (N+2)$ correlation matrix $\Sigma$ on all the Wiener processes $W^1,\dots,W^N, W^A,W^B$. 

Given one global permutation $\pi$, the permutations are $\pi^A$ and $\pi^B$ which take the available fuels in each zone, $S^{A,\pi^A_k}$ and $S^{B,\pi^B_l}$, from the set $S^{\pi_i}$ of fuels. 

As a consequence, the global set $\Vb_t$ of random variables can be rewritten as:
\beq
\Vb_t = \begin{pmatrix}
	\log S^1_t \\
	\log S^2_t \\
	\vdots \\
	\log S^N_t \\
	D^A_t \\
	D^B_t
\end{pmatrix}
\eeq
and is Gaussian conditionally to $\Vb_s$ for any $s <t$, that is, $\Vb_t|\Vb_s \sim \Nc(\mub(s,t)~;~\Sigmab(s,t))$, with $\mub(t,T)$ and $\Sigmab(t,T)$ detailed in appendix \ref{appendix_forward}.

\section{Derivatives pricing}
This section is dedicated to the computation of classical derivatives like forward prices, European call options, and geographical spread options to value the transmission rights from interconnection. In the following we ignore interest rates for notational simplicity.



%
 
 \subsection{Forward prices}
 

The forward price is set to be the expectation of the spot price under the pricing measure $\Qbb$:

\begin{equation*}
	F_t^*\left(T\right) = \mathbb{{E}}^\Qbb \left[\bar{P}^*\left(\Vb_T \right) | \mathcal{F}_t \right] = \Ebb_t^\Qbb \left[\bar{P}^* \left( \Vb_T\right) \right], \quad *=A,B 
\end{equation*}	

Using the partition defined by all permutations $\pi\in \Pi$ of the production costs, the power forward prices are then
\begin{equation}
\label{forwardAswitch}
\begin{split}
 F_t^A\left(T\right) = \sum_{\pi \in \Pi}\sum_{k=0}^{n_A} \sum_{l=0}^{n_B} ~
& \Ebb_t^\Qbb \left[ f^A(S^{A,\pi_{k}^A}_T,\bar{C}^A_T,D^A_T+\bar{E}) \unbb_{\Ac_{1,k,l}}(\Vb_T) \unbb_{\Sc^{\pi}}(\Vb_T)\right] + \\
&  \Ebb_t^\Qbb \left[f^A(S^{A,\pi_{k}^A}_T,\bar{C}^A_T,D^A_T+\ubar{E}) \unbb_{\Ac_{2,k,l}}(\Vb_T) \unbb_{\Sc^{\pi}}(\Vb_T)\right] +\\\\
&  \Ebb_t^\Qbb \left[f^{A,B}(\Vb_T)\unbb_{\Ac_{3,k,l}}(\Vb_T) \unbb_{\Sc^{\pi}}(\Vb_T) \right]
\end{split}
\end{equation}

and 

\begin{equation}
\label{forwardBswitch}
\begin{split}
F_t^B\left(T\right)  = 
\sum_{\pi \in \Pi}\sum_{k=0}^{n_A} \sum_{l=0}^{n_B} ~
&  \Ebb_t^\Qbb \left[f^B(S^{B,\pi_{l}^B}_T,\bar{C}^B_T,D^B_T-\bar{E}) \unbb_{\Ac_{1,k,l}}(\Vb_T) \unbb_{\Sc^{\pi}}(\Vb_T)\right] + \\
&  \Ebb_t^\Qbb \left[f^B(S^{B,\pi_{l}^B}_T,\bar{C}^B_T,D^B_T-\ubar{E}) \unbb_{\Ac_{2,k,l}}(\Vb_T) \unbb_{\Sc^{\pi}}(\Vb_T)\right] +\\\\
&  \Ebb_t^\Qbb \left[f^{A,B}(\Vb_T)\unbb_{\Ac_{3,k,l}}(\Vb_T) \unbb_{\Sc^{\pi}}(\Vb_T)\right]
\end{split}
\end{equation}

All of the terms in these expressions are of the form $\Ebb_t \left[e^{\lambda^T \Vb_T + \eta}\unbb_{\ab \le M\Vb_T \le \bb} \right]$. With the Gaussian assumption on $\Vb_t$ and Lemma \ref{lemma_laplace} of appendix \ref{appendix_forward}, these terms can be calculated by the computation of a probability of linear inequality constraints of the form $\Pbb(\ab \le \Mb \tilde{\Vb}_T \le \bb )$, under a multivariate Gaussian variable $\tilde{\Vb}_T$. A great deal of literature exists on the accurate numerical computation of the multivariate Gaussian distribution function. We refer to \cite{genzbook} for a recent review and the study of the specific linear inequality constraints in chapter 5.

\subsection{Transmission rights valuation}
\label{sec_transmission_rights}
In Europe, transmission rights are mainly physical and auctioned at yearly and monthly delivery. These contracts give the owner the right to transfer electricity  across the interconnection in the specified direction by the contract at any hour of the year or month. If the owner does not nominate its physical right and when markets are coupled, the owner receives financial compensation equal to the spot spread ("Use It or Sell It" condition). Physical Transmission Rights (PTR) combined with the "Use It or Sell It" condition are equivalent to a Financial Transmission Rights (FTR).\\

The valuation of transmission rights is equal to geographical spread options which correspond to the interconnection \cite{MahringerFussProkopczuk}, that is, the valuation at time $t$ of the option of payoff $(\bar{P}^B(\Vb_T)- \bar{P}^A(\Vb_T))^+ + (\bar{P}^A(\Vb_T)- \bar{P}^B(\Vb_T))^+$ at the maturity date $T>t$. By using the formulations \eqref{spotAswitch} and \eqref{spotBswitch} of the spot prices, we have:
\beqnx
 (\bar{P}^B(\Vb_T)- \bar{P}^A(\Vb_T))^+  =
\sum_{\pi ,k,l} && \hspace*{-6mm}
\Big [f^B(S^{B,\pi^B_l}_T,\bar{C}^B_T,D^B_T-\bar{E})-  \\
& & f^A(S^{A,\pi^A_k}_T,\bar{C}^A_T,D^A_T+\bar{E}) \Big ] \unbb_{\Ac_{1,k,l}\cap \Sc^{\pi}}(\Vb_t)  \\
(\bar{P}^A(\Vb_T)- \bar{P}^B(\Vb_T))^+  =
\sum_{\pi ,k,l} &&
\hspace*{-6mm}\Big [f^A(S^{A,\pi^A_k}_T,\bar{C}^A_T,D^A_T+\ubar{E})- \\
& & f^B(S^{B,\pi^B_l}_T,\bar{C}^B_T,D^B_T-\ubar{E}) \Big ] \unbb_{\Ac_{2,k,l}\cap \Sc^{\pi}}(\Vb_t)  \\
\eeqnx  
Therefore, the pricing of the geographical spread options only requires
some terms already used to valuate the forward prices.

\subsection{European Call options}
In this section, we focus on European call options. For example, we consider the valuation $\Cc_t$, at time $t$, of the option for payoff $(\bar{P}^A(\Vb_T)- K)^+$ at the maturity date $T>t$. The events are defined as:
\beqnx
\Bc_{1,k}& = & \left\{\omega \in \Omega: f^A(S^{A,\pi_{k}^A}_t,\bar{C}^A_t,D^A_t+\bar{E}) \ge K \right\} \\
\Bc_{2,k}& = & \left\{\omega \in \Omega: f^A(S^{A,\pi_{k}^A}_t,\bar{C}^A_t,D^A_t+\ubar{E}) \ge K \right\} \\
\Bc_{3,k}& = & \left\{\omega \in \Omega: f^{A,B}(\Vb_t) \ge K \right\}
\eeqnx
for $k=1,\dots,n_*$. These events are used to add the term $\unbb_{\bar{P}^A(\Vb_T) \ge K}$ to the valuation. We have:
\beqnx
\Cc_t  = \sum_{\pi,k,l} && \hspace*{-6mm} 
\Ebb_t^\Qbb \left[ 
f^A(S^{A,\pi_{k}^A}_t,\bar{C}^A_t,D^A_t+\bar{E}) \unbb_{\Ac_{1,k,l}\cap \Sc^{\pi}\cap\Bc_{1,k}}(\Vb_t)  \right] - K \Qbb\left( \Ac_{1,k,l}\cap \Sc^{\pi}\cap\Bc_{1,k} \right) \\
 + &&\hspace*{-6mm}\Ebb_t^\Qbb \left[ f^A(S^{A,\pi_{k}^A}_t,\bar{C}^A_t,D^A_t+\ubar{E}) \unbb_{\Ac_{2,k,l}\cap \Sc^{\pi}\cap\Bc_{2,k}}(\Vb_t) \right] - K \Qbb\left( \Ac_{2,k,l}\cap \Sc^{\pi}\cap\Bc_{2,k} \right)  \\
 +&&\hspace*{-6mm}\Ebb_t^\Qbb \left[ f^{A,B}(\Vb_t)\unbb_{\Ac_{3,k,l}\cap \Sc^{\pi}\cap\Bc_{3,k}}(\Vb_t) \right]  - K \Qbb\left( \Ac_{3,k,l}\cap \Sc^{\pi}\cap\Bc_{3,k} \right)
\eeqnx

The events $\Bc_{i,k}$ can be expressed by linear inequality constraints, as well as $\Ac_{i,k,l}$ and $\Sc^{\pi}$, on $\Vb_T$. We can then use the same tools previously used in the computation of of all the terms in the previous equation.
\section{Numerical illustrations}
In this section we give an illustration of the results obtained with the proposed model. In the following we focus on only one future date $T$. Therefore, we do not need to specify any seasonality in the demand and the parameters of the Ornstein-Uhlenbeck processes. The objective of this section is to first present an analysis on the reconstructed forward prices and the options for one future date $T$, and especially to understand the impact of some parameters on the prices. As we will see, the behavior of the derivative prices is quite comprehensible and depends on the initial conditions.
\subsection{Description of the example}
\label{sec_description_example}
\begin{figure}[!h]
		\hspace*{-6mm}
		\begin{tabular}{cc}
			Market A & Market B \\
			\includegraphics[scale=0.13]{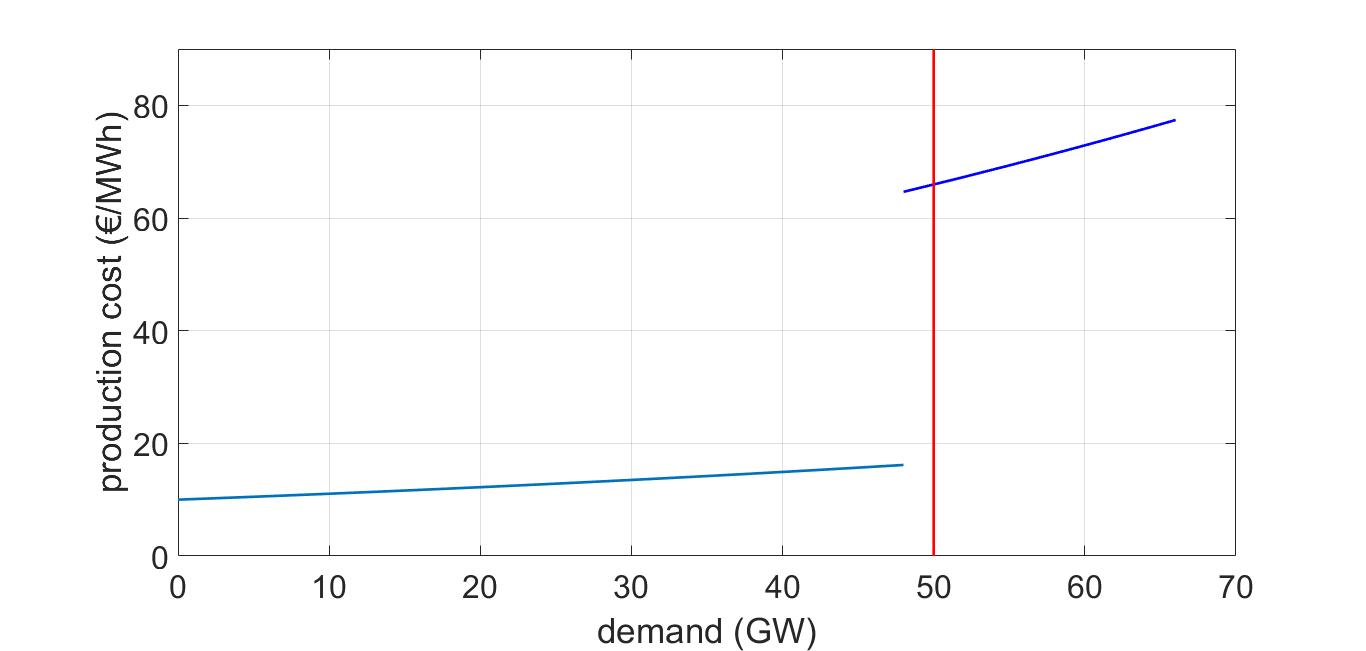} 	
			&
			\includegraphics[scale=0.13]{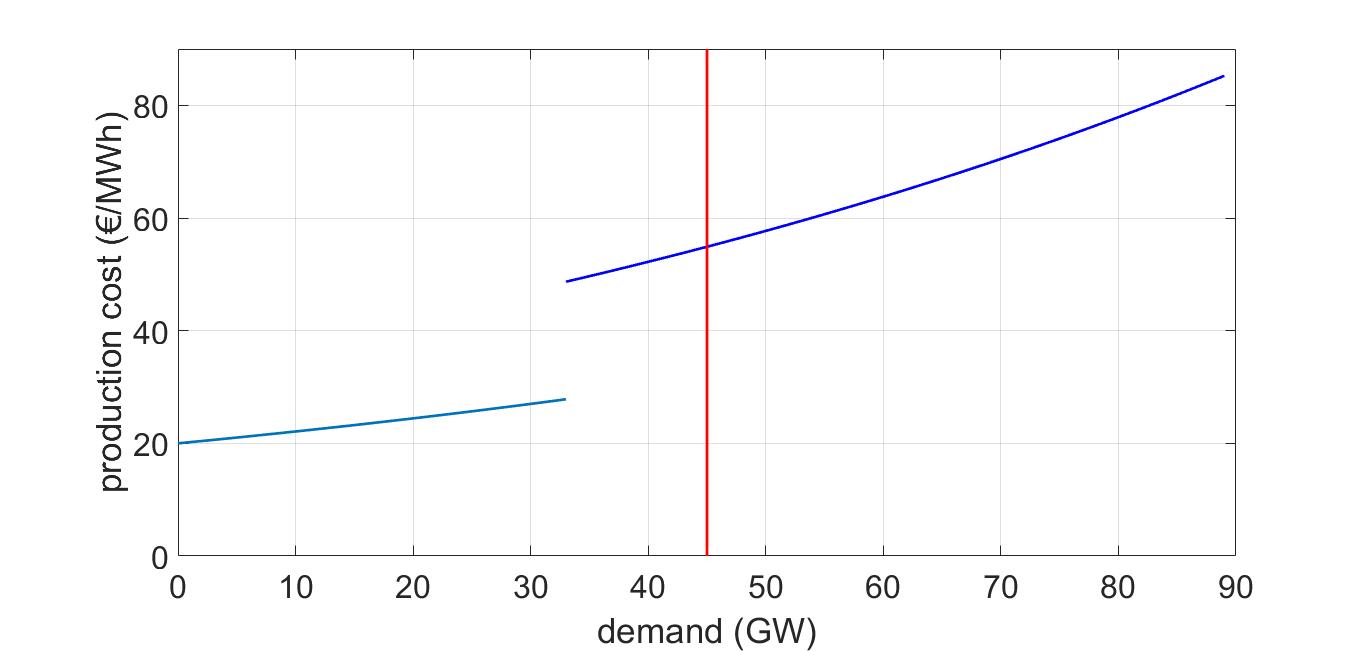} 	
						
		\end{tabular} 
		\caption{\label{production_cost} Average production costs of the two markets against mean demand level for market A (left) and market B (right), with $\beta^A=\beta^B=-0.01$,  $\alpha^A = 0.56$ and $\alpha^B = 0.89$ }
	
\end{figure}

We consider two neighboring markets which both have two different production technologies. We consider a Market A with a large cheap capacity and few peak plants (i.e., more expensive production technology). Market B is also composed of two technologies but its production costs are, on average, in between those of Market A. The markets' characteristics are described in table \ref{table_data}.
\begin{table}[h!]
	\renewcommand{\arraystretch}{1.3}
	\setlength{\tabcolsep}{0.5cm}
	\begin{center}
	\begin{tabular}{|c|c||c|c|}
		\hline 
		\multicolumn{2}{|c||}{Market A} &\multicolumn{2}{|c|}{Market B} \\
		\hline
		$C^{A,1}_{T}$ & 48GW & 	$C^{B,1}_{T}$ & 33GW \\
		\hline 
		$C^{A,2}_{T}$ & 18GW & 	$C^{B,2}_{T}$ & 56GW\\
			\hline 	\hline
		$	\Ebb_t[D^A_T]$ & 50GW & $\Ebb_t[D^B_T]$ & 45GW \\
		\hline 	\hline
		$\Ebb_t[\log S^{A,1}_{T}]$ & $\log 10$ & $\Ebb_t[\log  S^{B,1}_{T}]$ & $\log 20$\\
		\hline
		$\Ebb_t[\log S^{A,2}_{T}]$ & $\log 40$ & $\Ebb_t[\log S^{B,2}_{T}]$ & $\log 35$ \\
		\hline		
	\end{tabular}
	\end{center}
	\caption{Characteristics of the markets}
	\label{table_data}
\end{table} 
The parameters of function $f$ are set to $\beta^A=\beta^B=-0.01$,  $\alpha^A = 0.56$ and $\alpha^B = 0.89$ and Figure \ref{production_cost} illustrates the average production costs (offer curve) of the two markets.

If we consider deterministic demand levels and fuel costs, we trivially obtain the following results.
\bit
\item The global merit order is $(S^{A,1}_T,S^{B,1}_T,S^{B,2}_T,S^{A,2}_T)$.
\item Without any interconnection, Market A and Market B need to operate their two types of production assets in order to satisfy their corresponding level of demand.
\item When the interconnection is higher than 2GW, the technology $S^{A,2}_T$ is no longer operated and replaced by the added demand in Market B, that by technology $S^{B,2}_T$ satisfied.
\item The level of the interconnection has an impact on the marginal cost of Market A, whereas the one in Market B is always related to the technology $S^{B,2}_T$, whatever the level of the interconnection.
\eit
In the case where we add random levels of demand, we expect the price behaviour to change but to remain close to the ones previously described. In the case where we add random levels of fuel prices, we expect greater impacts. In particular, the merit order might change. For example, if $S^{B,2}_T$ becomes higher than $S^{A,2}_T$, then we expect a behaviour change to happen when the interconnection reaches 12GW. \\

In the next paragraphs, we give results from the following assumptions:
 \bit
  \item no correlation between fuels nor demands
  \item two different values for the demand's variance: low level (0.5) and high level (5)
  \item two different values for the fuel prices' volatility : low level (1\%) and high level (10\%)
  \eit

In the low fuel volatility case, the merit order has only 18\% probability to  change, that is the probability to observe switches between technologies is low. The increase of interconnection capacity will therefore decrease the occurrence of requiring the most expensive production assets ($S^{A,2}$) to satisfy the global demand. Next results are mostly explained by this fact and also by the distance, in average, between the demand levels and the discontinuity points of the offer curves.


\subsection{Forward pricing}
\label{sec_forward_prices}
\begin{figure}[!h]
	\begin{center}
		
		\begin{tabular}{ll}
			{\tiny low fuel cost volatility, low demand variance} & {\tiny low fuel cost volatility, high demand variance}
			\\
			\includegraphics[scale=0.125]{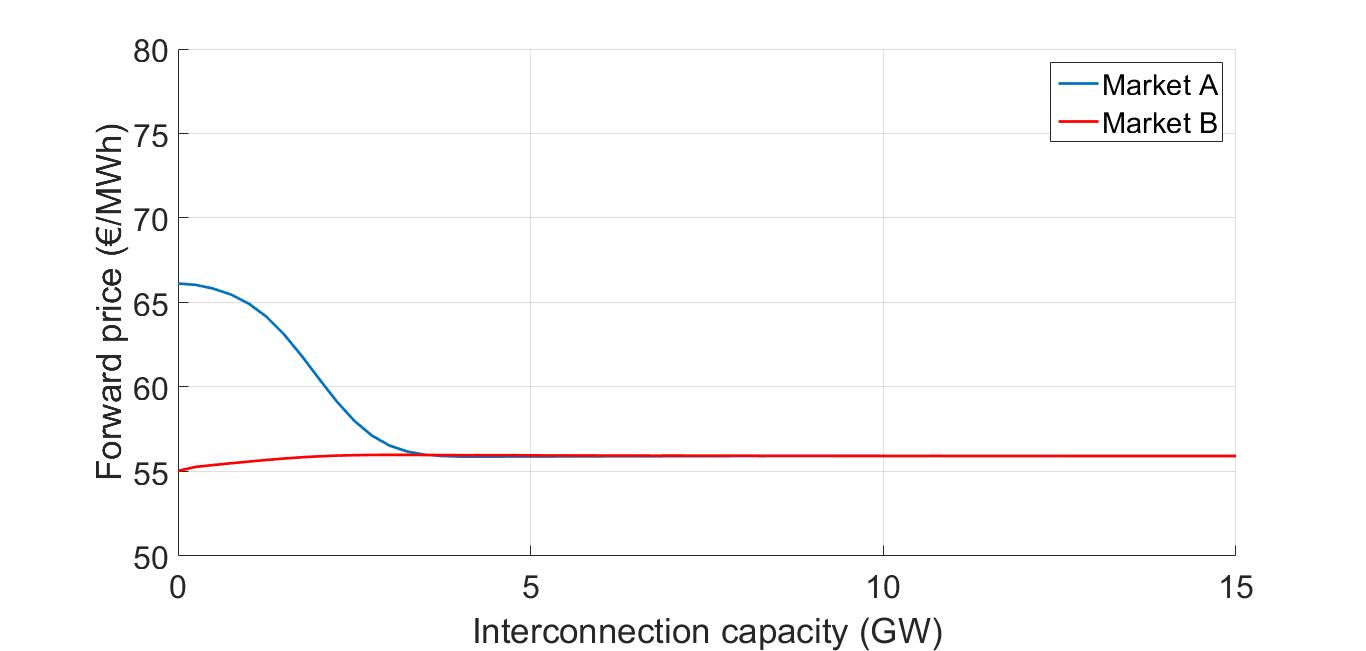} 	
			&
			\includegraphics[scale=0.125]{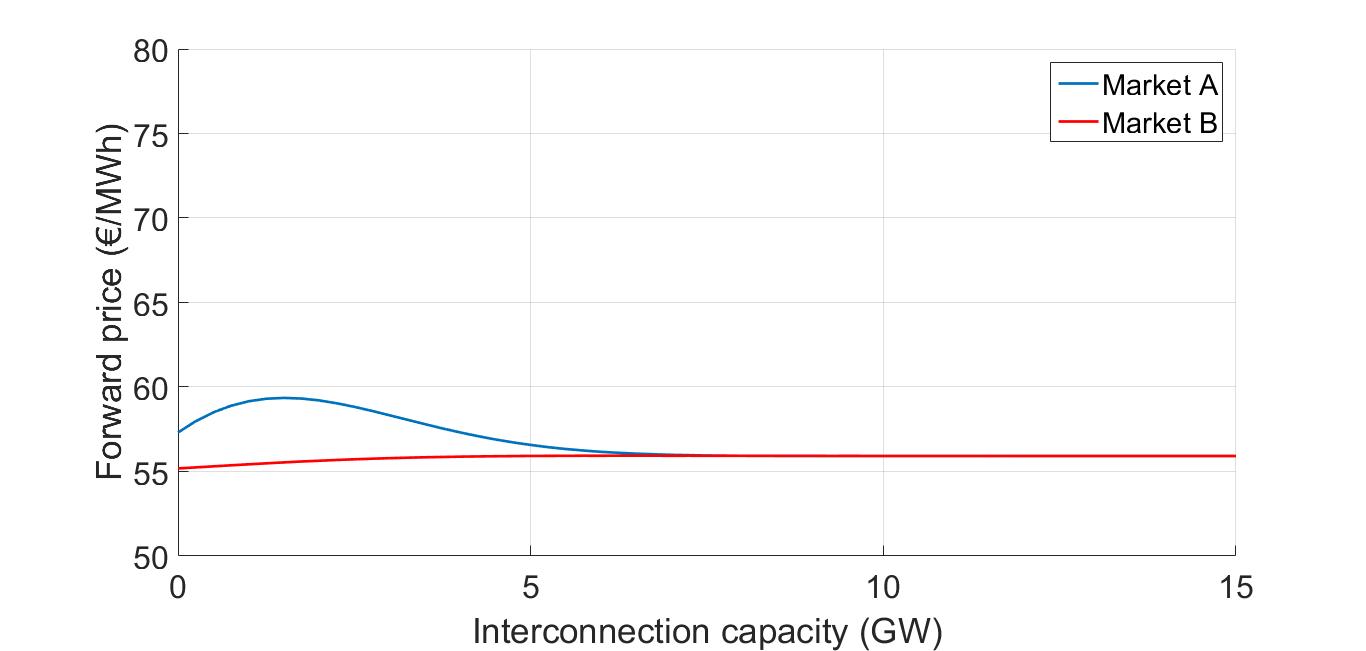}	 	
			\\ \\
			{\tiny high fuel cost volatility, low demand variance} & {\tiny high fuel cost volatility, high demand variance}
			\\
			\includegraphics[scale=0.125]{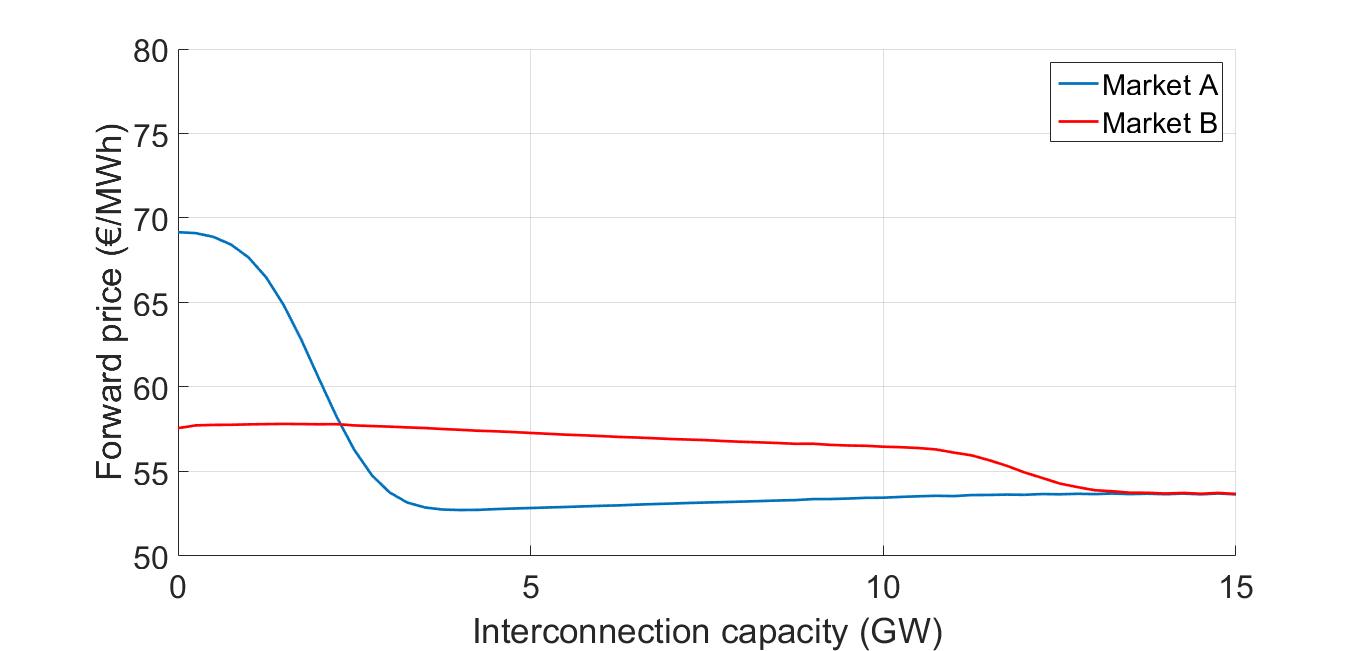} 	
			&
			\includegraphics[scale=0.125]{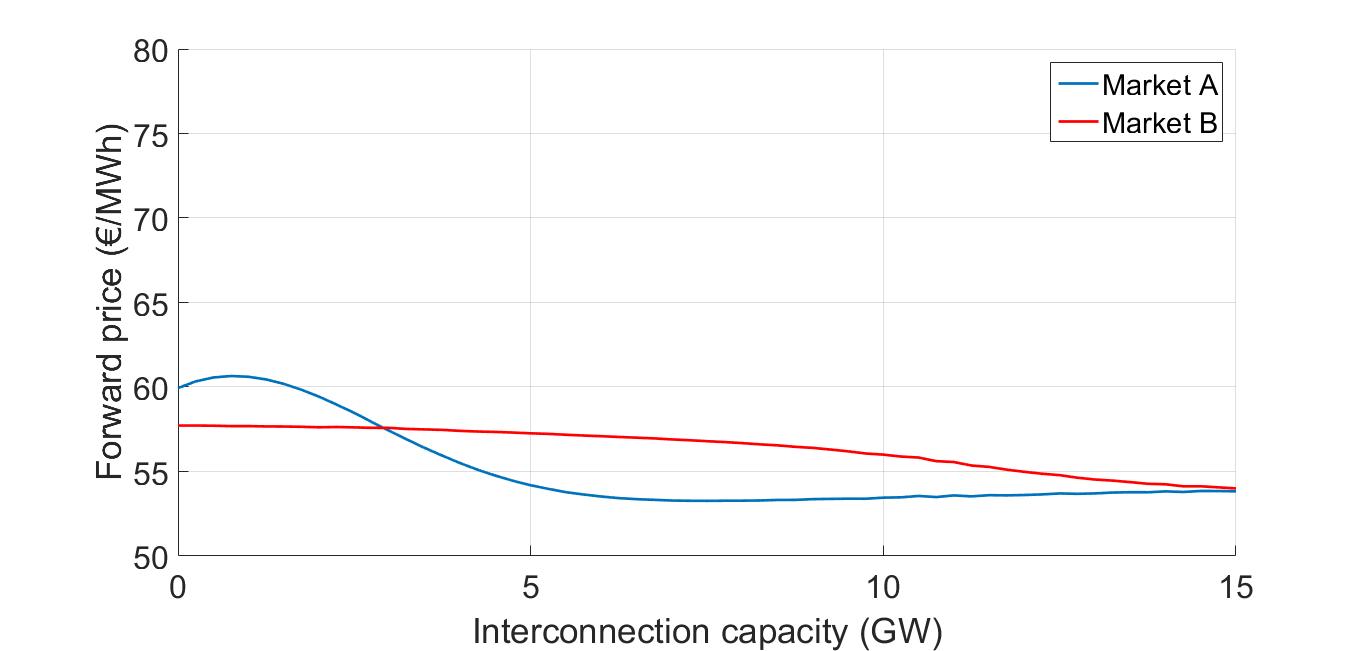} 	
			
		\end{tabular} 
		\caption{\label{forward} Forward prices of the two markets against interconnection capacities in four cases: low (left) and high (right) variances in demand, low (top) and high (bottom) volatilities in fuel prices.}
	\end{center}
	
\end{figure}

\begin{figure}[!h]
	\begin{center}
		
		\begin{tabular}{cc}
			{\tiny low fuel cost volatility, low demand variance} & {\tiny low fuel cost volatility, high demand variance}
			\\
			\includegraphics[scale=0.125]{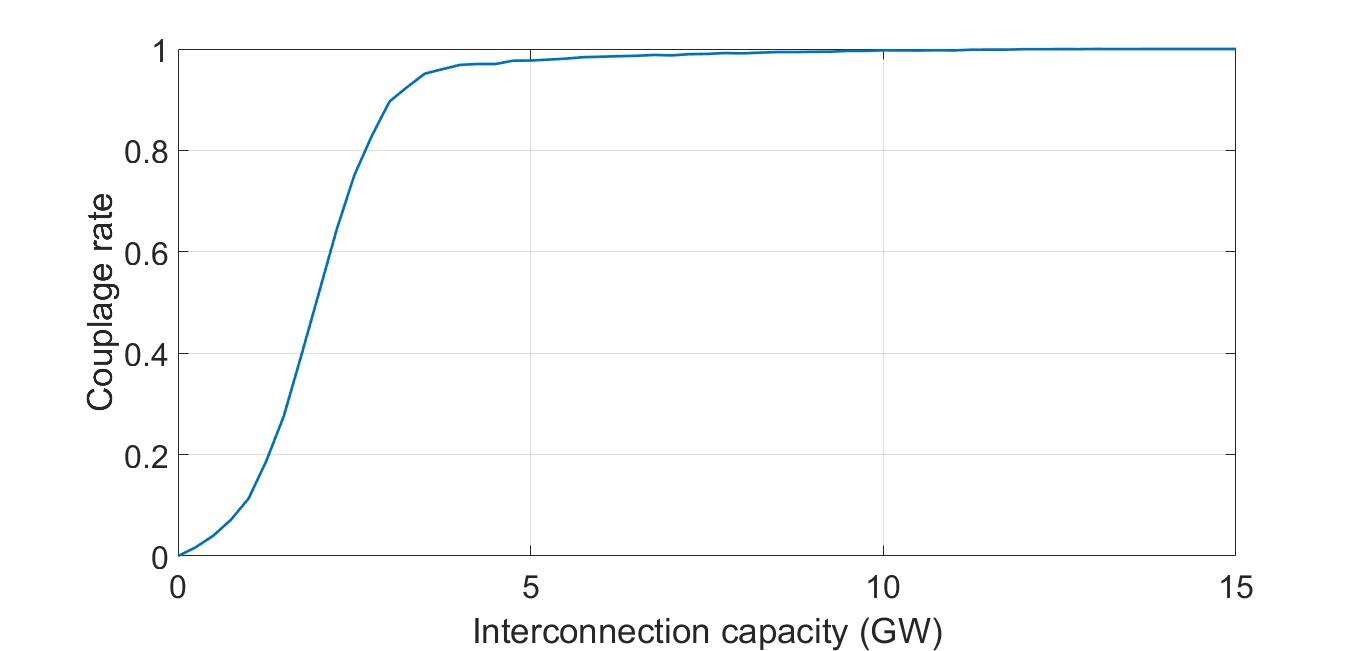} 	
			&
			\includegraphics[scale=0.125]{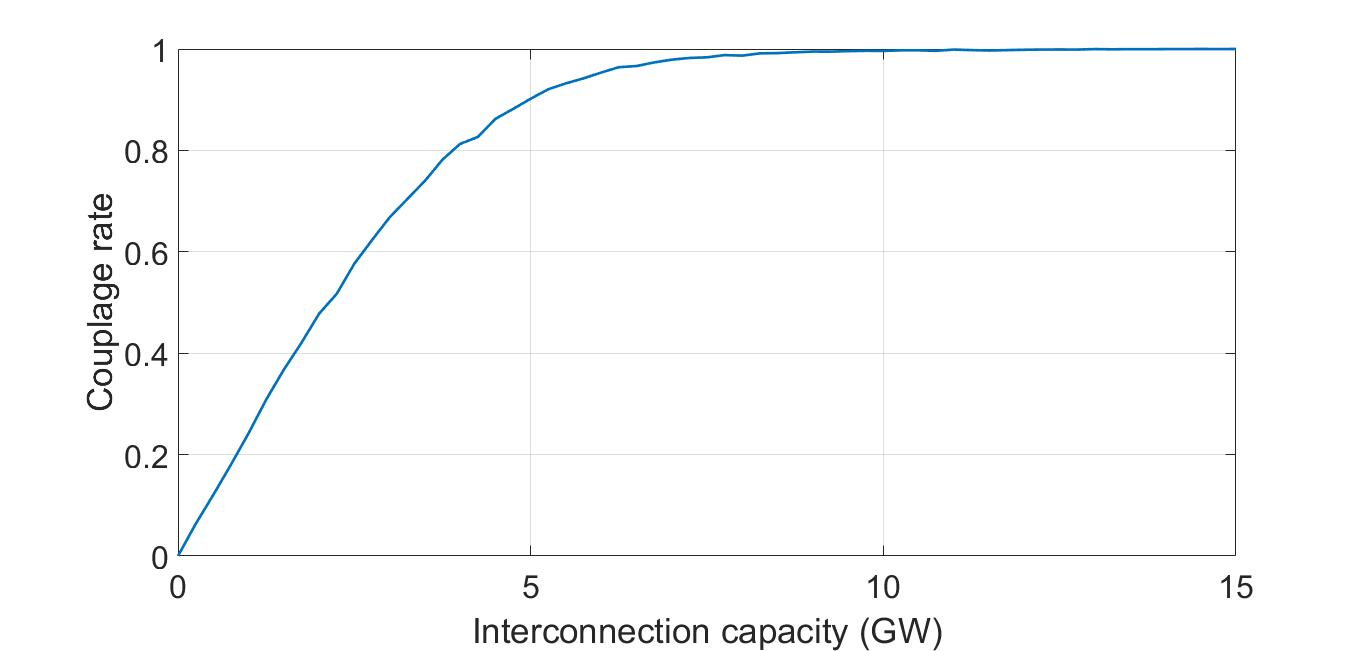}	 	
			\\ \\
			{\tiny high fuel cost volatility, low demand variance} & {\tiny high fuel cost volatility, high demand variance}
			\\
			\includegraphics[scale=0.125]{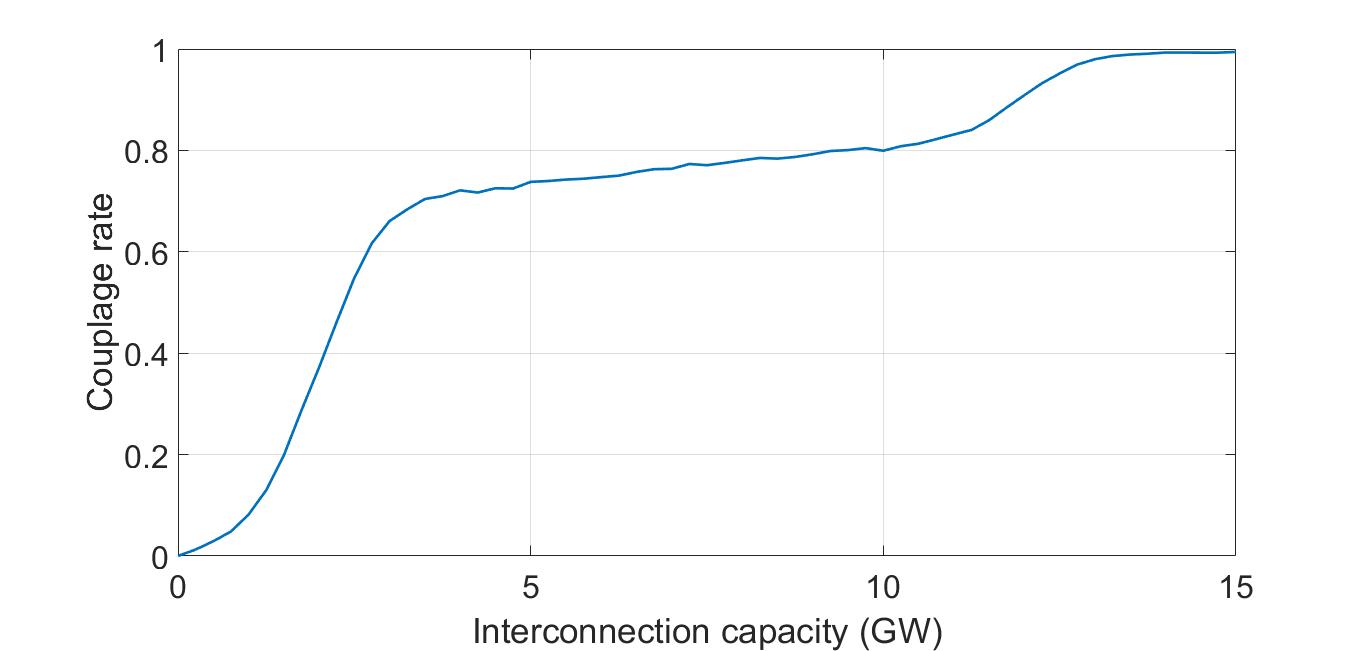} 	
			&
			\includegraphics[scale=0.125]{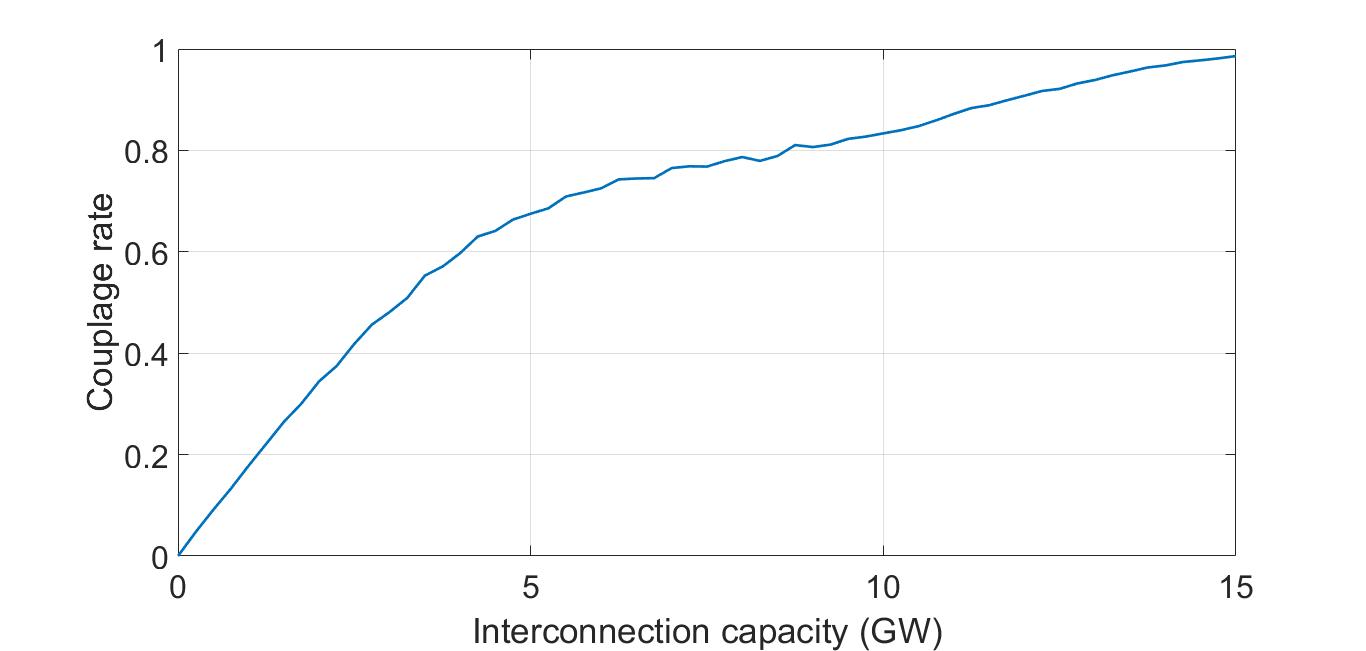} 	
			
		\end{tabular} 
		\caption{\label{coupling_rate} Coupling rate of the two markets against interconnection capacities in four cases: low (left) and high (right) variances in demand, low (top) and high (bottom) volatilities in fuel prices.}
	\end{center}
	
\end{figure}

In this subsection, we analyse the impact of the interconnection capacity $\bar E=-\ubar E$ on the forward prices (figure \ref{forward}) and on the coupling rate (figure \ref{coupling_rate}), that is, the percentage of time during which the two spot prices are strictly equal.

As expected, the coupling rate increases with the interconnection capacity. A 100\% coupling rate corresponds to when the interconnection capacity is high enough so that the two markets are one. In parallel, the forward prices converge to an identical value for the two markets. In every case, due to the supposed mean levels of demands and capacities, the forward prices converge to a price that is mainly related to the fuel cost $S^{B,2}$. However, behaviours of the forward prices and the coupling rate against the interconnection capacity $E$ drastically change with the assumptions of the demands' variance and the fuel costs' volatility. 

When both the demand variance and fuel cost volatilities are low, we retrieve the same results as for the coupling rate, the quasi perfect forward price convergence appears for a value of the interconnection capacity that is higher than 2GW due to the (even) small uncertainty on demands and fuel costs. 

When the demands' variance increases, we observe a slower convergence in the forward prices, which is trivially explained by an increasing probability that the spread between prices $P^A$ and $P^B$ is high, and then an increasing probability of the events $\Ac_1$ and $\Ac_2$. In these cases one needs a higher interconnection capacity to ensure the price convergence. 
\\
We can also observe different behaviors in the forward prices with respect to the demand variances, especially the forward price in Market A for low interconnection capacities: \\
\begin{itemize}
\item When there is no interconnection, the forward price in Market A is lower for high demand variances. This price is explained by the fact that a high demand variance increases the probability of $S^{A,1}_T$ being marginal, whereas for the low demand variance, the probability of $S^{A,2}$ being marginal is close to one. Due to the discontinuity in the offer curve, this probability leads to a lower forward price.
\item When the interconnection increases, the probability of $S^{A,1}$ being marginal decreases due to the fact that it has a high probability of being the cheapest fuel (it has a high probability of being completely used to satisfy the global demand), and then the forward price in Market A increases. 
\item When the interconnection capacity is high enough, the forward price in Market A decreases because the probability of $S^{A,2}$ being marginal decreases (it is replaced by the available cheaper fuel in Market B).

\end{itemize}

Concerning the behaviour changes with respect to the fuel cost volatility, we have two main observations. \\
\bit
\item The price convergence is weaker when the fuel cost volatility is high. This effect is due to the log-normality of the fuel costs, which decreases the mean value when the volatility increases. This is then an artificial effect due to the simple example shown in this section. \\
\item The results show a "crossing effect" which makes the forward price in Market A lower than in Market B for middle values of interconnection capacities. This effect highlights the impact of fuel cost volatility in the forward prices. \\
When the fuel cost volatilities are low, there is a weak probability of switches in the global merit order. In this example the two most probable events are: i) the fuel $S^{A,2}_T$ is the most expensive fuel and ii) the fuel $S^{B,2}_T$ is the most expensive fuel. Due to the average levels of demand, when the interconnection capacity is in middle values, it is high enough to substitute $S^{A,2}_T$ with the cheaper fuels in Market B when $S^{A,2}_T$ is the most expensive fuel. At the same time, the interconnection capacity is not high enough to replace $S^{B,2}$ with cheaper fuels in Market A when $S^{B,2}$ is the most expensive fuel. Therefore, in the two most probable events, the price in Market A is lower or equal to the one in Market B. When the interconnection is high enough (higher than 14GW or 15GW for low or high variance in demand, respectively), the most expensive fuel ($S^{A,2}$ or $S^{B,2}$) is always replaced by cheaper fuels.  
\eit

\subsection{Transmission rights pricing}
\label{sec_interco_option}
\begin{figure}[!h]
	\begin{center}
		\hspace*{-5mm}
		\begin{tabular}{cc}
			{\tiny low fuel cost volatility, low demand variance} & {\tiny low fuel cost volatility, high demand variance}
			\\
			\includegraphics[scale=0.105]{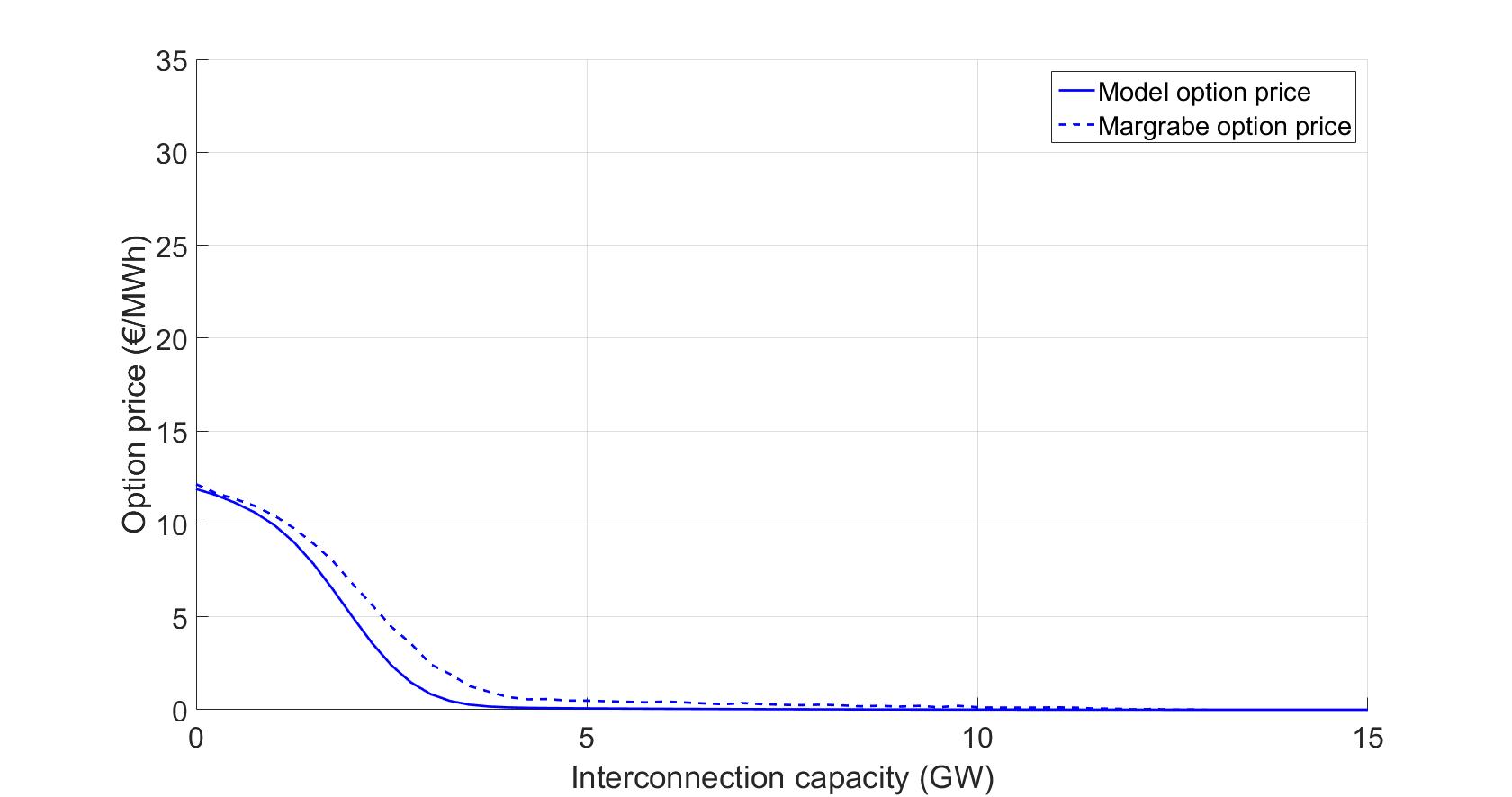} 	
			&
			\includegraphics[scale=0.105]{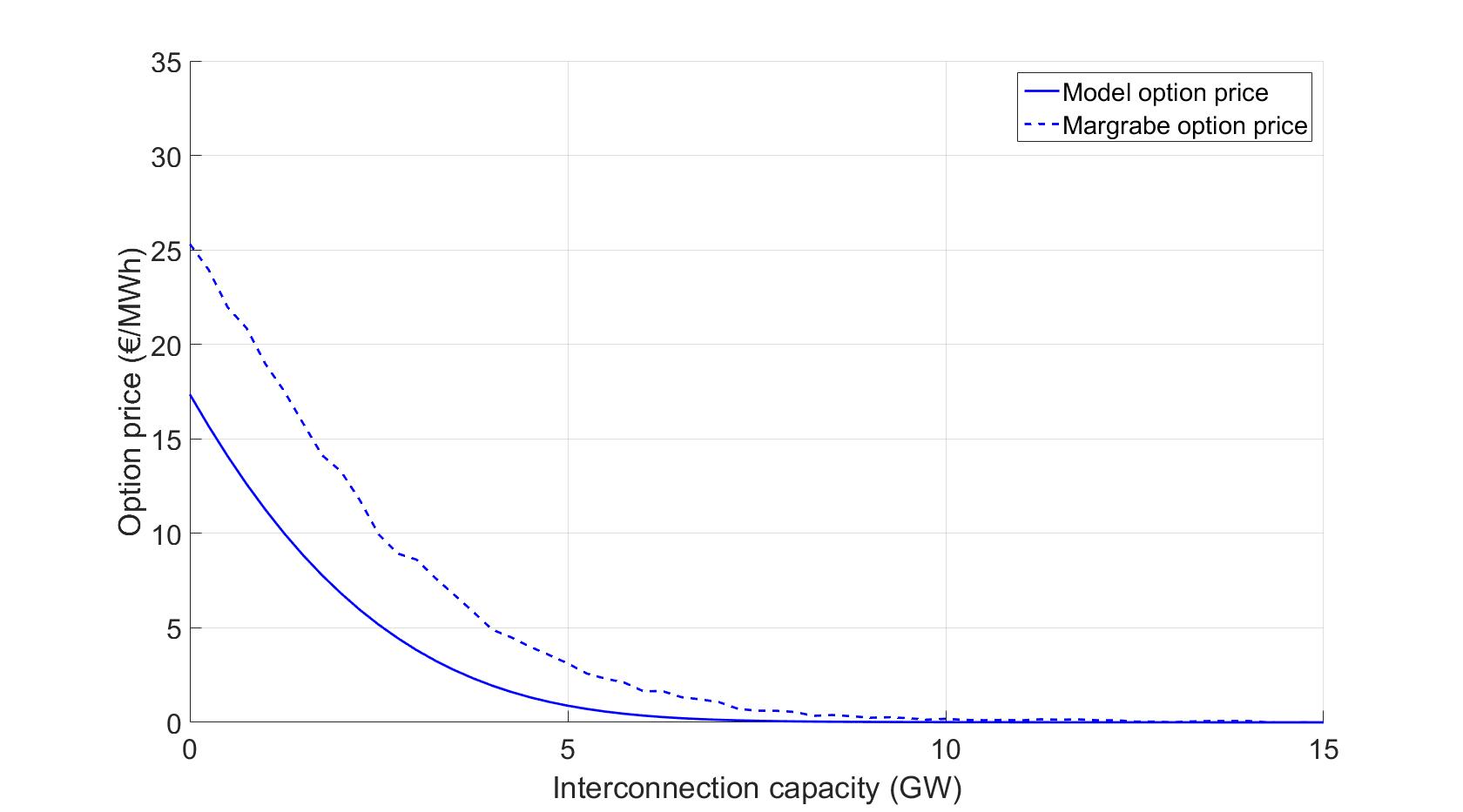}	 	
			\\ \\
			{\tiny high fuel cost volatility, low demand variance} & {\tiny high fuel cost volatility, high demand variance}
			\\
			\includegraphics[scale=0.105]{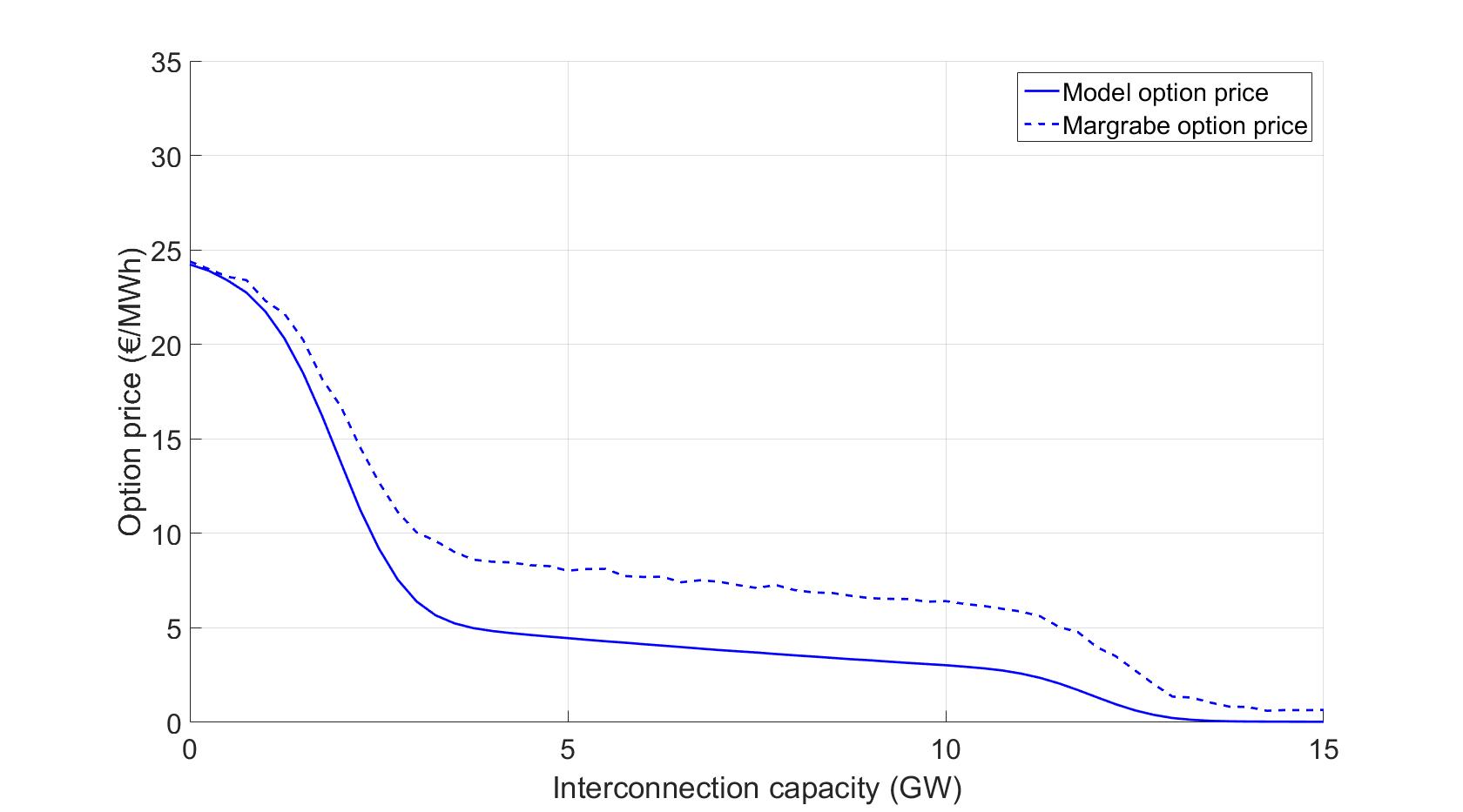} 	
			&
			\includegraphics[scale=0.105]{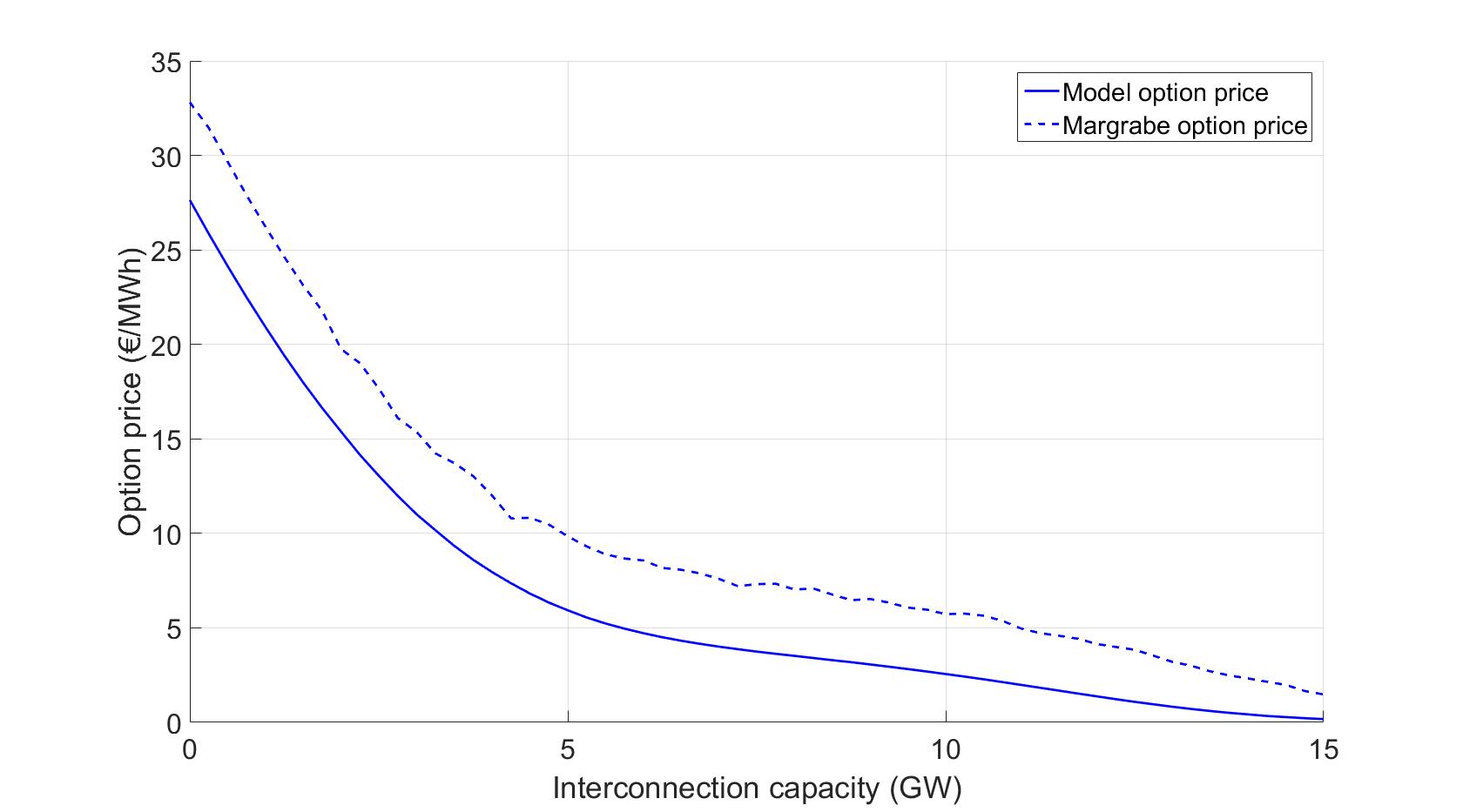} 	
			
		\end{tabular} 
		\caption{\label{PTR} Transmission right pricing against the interconnection capacity with analytical formulas (solid lines) and the Margrabe formula (dashed lines) in four cases: low (left) and high (right) variances in demand, low (top) and high (bottom) volatilities in fuel prices.}
	\end{center}
	
\end{figure}

In this section we are interested in the pricing of options for transmission rights with respect to the interconnection capacity and with respect to level of uncertainty for the demand and fuel costs. Specifically, we represent the option prices of payoff $(\bar{P}^B(\Vb_T)- \bar{P}^A(\Vb_T))^+ + (\bar{P}^A(\Vb_T)- \bar{P}^B(\Vb_T))^+ $ by allowing the owner to use the interconnection in both directions.  Figure \ref{PTR} shows the transmission right prices with respect to the transmission capacity $E$  for any cases of demand variance and fuel cost volatilities. The prices are obtained by using the same parameter values as in the previous section. We can also observe that the behaviour of the transmission right’s price is similar to the coupling rate’s behaviour (figure \ref{coupling_rate}). This similarity is logical due to the link between the option value and the probability of convergence in the power spot prices, which leads to a zero option value when the interconnection capacity is high enough. Thus, the power spot prices are almost surely identical in both markets. \\

In addition, we compare these prices to the approximated option prices by using the Margrabe formula. Figure \ref{PTR} presents the results of the structural model’s option prices (plain lines) compared to the Margrabe formula’s results (dotted lines). In order to get the Margrabe’s option prices, we estimate the spot price volatilities and correlations (needed in the Margrabe formula) on spot prices simulated by the proposed model. In doing so, the differences in transmission right prices between the two models are only linked to the valuation formula and not to volatility or a poor evaluation of the correlation. \\
As a global remark, we can observe, as in \cite{MahringerFussProkopczuk}, that the Margrabe formula leads to an overestimation of the transmission right prices. The observed difference can be very high: for many sets of parameters and capacity values, the Margrabe approximation can give prices that are twice the prices given by the structural coupled model. Therefore, in practice to apply the Margrabe formula to price the transmission rights can lead to a poor bet on long-term explicit auctions for interconnection capacities. The difference between the two prices measures the error made by simplifying the coupled price dynamics to the correlated price dynamics. Only in the case of small interconnection capacities and small demand variances are the prices from the model and Margrabe formula identical. \footnote{The case of no interconnection seems contradictory to a positive value of the transmission rights option. But the pricing formulas give prices in \euro /MWh and implicitly suppose that the owner can always arbitrate between the two markets.} In this case, only the fuel costs are really random. And the prices in Markets A and B are, with a high probability, given by the fuel costs $S^{A,2}$ and $S^{B,2}$ which are log-normally distributed. Therefore, the power spot prices in Markets A and B are quasi log-normally distributed and the structural model’s price then corresponds to the Margrabe formula. But when the demands' variance is high, the power spot prices are not log-normally distributed for the structural model and the price difference then appears even with no interconnection. \\

\begin{figure}[!h]

	\begin{center}
		\hspace*{-5mm}
		\begin{tabular}{cc}
			{\tiny $E$=0} & {\tiny $E$=4}
			\\
			\includegraphics[scale=0.2]{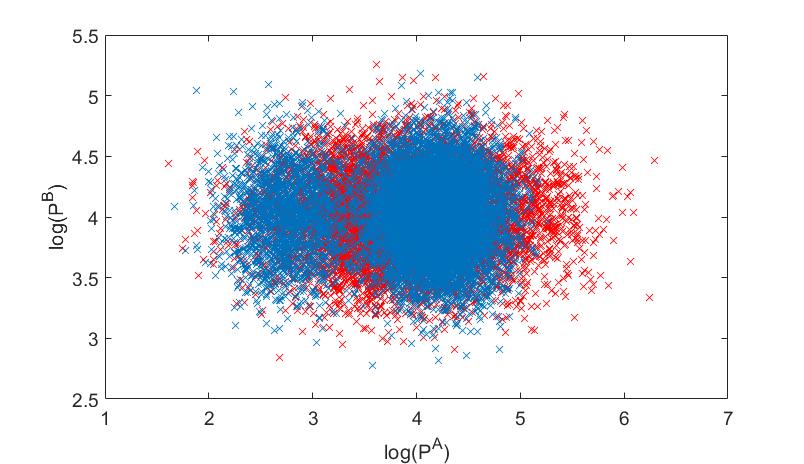} 	
			&
			\includegraphics[scale=0.2]{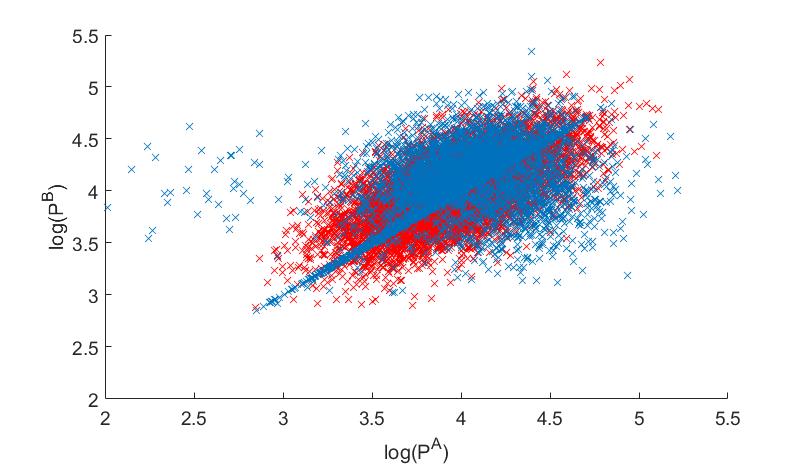}	 	
		\end{tabular} 
		\caption{\label{SpotSimulation} Log spot price $\bar{P}^A $ against log spot price $\bar{P}^B $ for $E=0$ (left) and $E=4$ (right) simulations for the proposed structural model in blue and the corresponding log-normal approximation in red, with high fuel cost volatility and high demand variance }
	\end{center}
	
\end{figure}

 The fact that the Margrabe formula gives higher prices compared to the proposed model can be explained by the fact that the Margrabe approach neglects the coupling mechanism. Indeed, with the market coupling mechanism, the prices’ equality occurs with a high probability that lowers the value of the interconnection transmission rights compare to the same markets with no coupling mechanism. To illustrate this effect, figure \ref{SpotSimulation} presents an example of spot prices in Markets A and B for the structural coupled model (blue dots) and for the approximated log-normal distribution (used in the Margrabe approximation) in red dots. This figure shows that our coupled model puts a high probability of occurrence on identical prices in both markets (i.e., high concentration of simulations on the main diagonal), which is not the case in the log-normal model when $E>0$.    \\

%
%
%
%
\section{Conclusion}
To cope with the large development of coupling in European electricity  markets, we propose in this paper a model which represents two interconnected power markets. This model is important when a portfolio is exposed to the spread between two neighbouring countries or to participation in long-term auctions to get explicit interconnection capacities. To benefit from higher liquidity in neighbouring markets is also a motivation to the portfolio manager to put in place  proxy-hedging strategies. And in order to efficiently set up these strategies, the coupling mechanism between markets needs to be represented. \\

In addition, a deep transformation in the electricity production mix  is happening. Renewables’ capacities are developing, and thermal plants are being decommissioned. These phenomena lead to changes in the characteristics of electricity prices and  make electricity price models based on a statistical approach difficult to calibrate and inefficient in representing the future characteristics of prices. This is why, the price model we propose is based on the structural approach which represents electricity as a by-product of renewables and thermal generation that are needed to satisfy demand. \\

We show how to compute the power spot prices of both markets with this model, as well as classical derivatives like forward prices, European options, and transmission right options. The model does all with analytical formulas, which is essential from a practitioner's point of view. We also show an illustration of the results obtained from the model and especially the impacts on derivative prices for the interconnection capacity and for the levels of volatilities. This illustration allows us to fully understand all of the effects and shows how the proposed model is able to deal with pricing. Especially, numerical examples illustrate how different the pricing of transmission rights is  between the proposed model, which explicitly considers the coupling mechanism, versus the Margrabe approach. \\

Our future work will focus on the calibration of the model to observed spot prices, for example, in the French and German markets. Further, the restriction to only two interconnected markets does not represent all behaviour in spot prices well but we expect to capture the most important one. 
Further, we will focus on the computation of forward prices for several maturities to build the real forward products (with a delivery period) and then compare these reconstructed prices to the forward prices observed in the markets, as done in \cite{Feron15} in the case of one market, and, eventually, proceed to the calibration of the model for observed forward prices.


\label{Bibliography}

\bibliographystyle{apalike}

\bibliography{biblio} 

\begin{thebibliography}{}

\bibitem[Aïd et~al., 2012]{aid2012structural}
Aïd, R., Campi, L., and Langren{\'e}, N. (2012).
\newblock A structural risk-neutral model for pricing and hedging power
  derivatives.
\newblock {\em Mathematical Finance}.

\bibitem[Barlow, 2002]{barlow2002diffusion}
Barlow, M.~T. (2002).
\newblock A diffusion model for electricity prices.
\newblock {\em Mathematical Finance}, 12(4):287--298.

\bibitem[Carmona and Coulon, 2012]{carmona2012survey}
Carmona, R. and Coulon, M. (2012).
\newblock A survey of commodity markets and structural models for electricity
  prices.
\newblock {\em Financial Engineering for Energy Asset Management and Hedging in
  Commodity Markets}.

\bibitem[Carmona et~al., 2013]{carmona2013}
Carmona, R., Coulon, M., and Schwarz, D. (2013).
\newblock Electricity price modeling and asset valuation: A multi-fuel
  structural approach.
\newblock {\em Matehmatics and Financial Economics}, 1(4):167--202.

\bibitem[Carr, 2012]{Carr}
Carr, G. (2012).
\newblock European power market: Cae's next challenges.
\newblock {\em Energy Risk}.

\bibitem[Commission, 2015]{EuropReport}
Commission, E. (2015).
\newblock Quaterly report on european electricity markets.
\newblock {\em DG Energy}, 8(3).

\bibitem[Coulon et~al., 2013]{Coulon2013}
Coulon, M., Power, W., and Sircar, R. (2013).
\newblock A model for hedging load and price risk in the texas electricity
  market.
\newblock {\em Energy Economics}, 40(4):976--988.

\bibitem[F\'eron and Daboussi, 2015]{Feron15}
F\'eron, O. and Daboussi, E. (2015).
\newblock {\em Calibration of electricity price models}, pages 183--210.
\newblock Fields institue communications edition.

\bibitem[Fuss et~al., 2014]{Fuss}
Fuss, R., Mahringer, S., and Prokopczuk, M. (2014).
\newblock Electricity spot and derivatives pricing when markets are
  interconnected.
\newblock {\em working paper}.

\bibitem[Genz and Bretz, 2009]{genzbook}
Genz, A. and Bretz, F. (2009).
\newblock {\em Computation of Multivariate Normal and t Probabilities}.
\newblock Lecture Notes in Statistics. Springer-Verlag, Heidelberg.

\bibitem[Kiesel and Kustermann, 2015]{kustermann}
Kiesel, R. and Kustermann, M. (2015).
\newblock Structural models for coupled electricity markets.
\newblock {\em working paper}.

\bibitem[Mahringer et~al., ]{MahringerFussProkopczuk}
Mahringer, S., Fuss, R., and Prokopczuk, M.
\newblock Electricity market coupling and the pricing of transmission rights:
  an option-based approach.
\newblock {\em working paper}.

\end{thebibliography}

\appendix
\section{Proposition proofs}
\label{sec_proof}
This section is devoted to prove the equivalences written in the Propositions \ref{prop_Ca} to \ref{prop_Cc}.
\subsection{Proof of Propositions \ref{prop_Ca}}
In this proposition we have $g(\Vb_t)=\sum_{i=0}^{k-1}C^{A,i}_t - D^A_t$. 

Assume $\Vb_t \in \Ac_{3,k,l}^A$. Therefore $\Vb_t \in \Mc_{k,l}$ and the optimal flow $E_t$ is such that $D^A_t + E_t$ is a point of discontinuity of $P^A$. This clearly implies that $E_t = g(\Vb_t)$ and $D^B_t - g(\Vb_t) \in I^{B,l}_t$.
Because $\Vb_t \in \Ac_3$, it follows that $\ubar{E}< g(\Vb_t) < \bar{E}$. 
There are two cases depending on the sign of $E_t$. 
\bit
\item if $E_t>0$ we are in the first case of Definition \ref{def_Et}, i.e. $P^A(\Sb_t^A,\Cb^A_t,D^A_t)\le P^B(\Sb_t^B,\Cb^B_t,D^B_t)$. Because $\Vb_t \in \Mc_{k,l}$ and by definition of $E_t$, this implies $P1$-3. Now suppose $f^A(S^{A,k}_t,\bar{C}^A_t,D^A_t + E_t) < f^B(S^{B,l}_t,\bar{C}^B_t,D^B_t - E_t)$ then there is $\varepsilon>0$ such that $E_t+\varepsilon<\bar{E}$ and
\beqx
f^A(S^{A,k}_t,\bar{C}^A_t,D^A_t + E_t+\varepsilon) < f^B(S^{B,l}_t,\bar{C}^B_t,D^B_t - E_t-\varepsilon)
\eeqx
which contradicts the definition of $E_t$.
\item if $E_t<0$ we are in the second case of Definition \ref{def_Et}, i.e. $P^A(\Sb_t^A,\Cb^A_t,D^A_t)\ge P^B(\Sb_t^B,\Cb^B_t,D^B_t)$. Because $\Vb_t \in \Mc_{k,l}$ and by definition of $E_t$, this implies $P1$-4. Now suppose $f^A(S^{A,k-1}_t,\bar{C}^A_t,D^A_t + E_t) > f^B(S^{B,l}_t,\bar{C}^B_t,D^B_t - E_t)$ then there is $\varepsilon>0$ such that $E_t-\varepsilon>\ubar{E}$ and
\beqx
f^A(S^{A,k-1}_t,\bar{C}^A_t,D^A_t + E_t-\varepsilon) > f^B(S^{B,l}_t,\bar{C}^B_t,D^B_t - E_t+\varepsilon)
\eeqx
wich contradicts the definition of $E_t$.
\eit

Now assume that $\Vb_t$ respects the inequalities $P1$-1 to $P1$-4 of proposition \ref{prop_Ca}. Note that if $E_t=g(\Vb_t)$ then we directly have $\Vb_t \in \Ac_{3,k,l}^A$. Therefore we only have to prove that $E_t=g(\Vb_t)$. By definition of $g(\Vb_t)$ and from $P1$-2 we conclude that
\beqnx 
P^A(\Sb_t^A,\Cb^A_t,D^A_t+g(\Vb_t)) & = & f^A(S^{A,k}_t,\bar{C}^A_t,D^A_t + g(\Vb_t)) \\
P^B(\Sb_t^B,\Cb^B_t,D^B_t-g(\Vb_t)) & = & f^B(S^{B,l}_t,\bar{C}^B_t,D^B_t - g(\Vb_t))
\eeqnx
\bit
\item if $g(\Vb_t)>0$ then $P^A(\Sb_t^A,\Cb^A_t,D^A_t)\le P^B(\Sb_t^B,\Cb^B_t,D^B_t)$ beacause $f^A$ and $f^B$ are increasing functions. From $P1$-3 we get that $\forall e<g(\Vb_t),~P^A(\Sb_t^A,\Cb^A_t,D^A_t+e)\le P^B(\Sb_t^B,\Cb^B_t,D^B_t-e) $, therefore $E_t \ge g(\Vb_t)$ by definition of $E_t$. And from $P1$-4 we have $E_t \le g(\Vb_t)$ which ends the fact that $E_t=g(\Vb_t)$. 
\item if $g(\Vb_t)<0$ then $P^A(\Sb_t^A,\Cb^A_t,D^A_t)\ge P^B(\Sb_t^B,\Cb^B_t,D^B_t)$ beacause $f^A$ and $f^B$ are increasing functions. From $P1$-4 we get that $\forall e>g(\Vb_t),~P^A(\Sb_t^A,\Cb^A_t,D^A_t+e)\ge P^B(\Sb_t^B,\Cb^B_t,D^B_t-e) $, therefore $E_t \le g(\Vb_t)$ by definition of $E_t$. And from $P1$-3 we have $E_t \ge g(\Vb_t)$ which ends the fact that $E_t=g(\Vb_t)$ 
\eit

It is trivial to show that if $\Vb_t \in \Ac_{3,k,l}^A$ the most expensive fuel needed to satisfy the global demand is $S^{B,l}_t$ and, because $g(\Vb_t)$ is not so that $D^B_t - g(\Vb_t)$ is a point of discontinuity of $P^B$, the common price is $f^{A,B}(\Vb_t)=f^B(S^{B,l}_t,\bar{C}^B_t,D^B_t - g(\Vb_t))$.
 \subsection{Proof of Proposition \ref{prop_Cb}}
 The proof of Proposition \ref{prop_Cb} is identical to the proof of Proposition \ref{prop_Ca}. One only needs to invert market $A$ and market $B$, and change the sign of $E_t$.

 \subsection{Proof of Proposition \ref{prop_Cc}}
 In this proposition we have set $g(\Vb_t)=\frac{\ln S_t^{A,k} -\ln S_t^{B,l} + \alpha^A - \alpha^B+  \beta^A(\bar{C}_t^{A}- D_t^A ) - \beta^B(\bar{C}_t^{B}- D_t^B)}{\beta^A + \beta^B}$. 
 
 Assume $\Vb_t \in \Ac_{3,k,l}^C$. Because $\Vb_t \in \Mc_{k,l}$ we have:
 \beqnx
 P^A(\Sb^A_t,\Cb^A_t,D^A_t + E_t) & = & f^A(S^{A,k}_t,\bar{C}^A_t,D^A_t + E_t) \\
 P^B(\Sb^B_t,\Cb^B_t,D^B_t - E_t) & = & f^B(S^{B,l}_t,\bar{C}^B_t,D^B_t - E_t) 
 \eeqnx
 Because $E_t$ is such that $D^A_t + E_t$ and $D^B_t - E_t$ are not discontinuity points of $P^A$ and $P^B$ respectively, the optimal flow $E_t$ must be such that $f^A(S^{A,k}_t,\bar{C}^A_t,D^A_t + E_t) = f^B(S^{B,l}_t,\bar{C}^B_t,D^B_t - E_t) $. Therefore $E_t=g(\Vb_t)$ and then $\Vb_t$ respects $P3$-1 to $P3$-3.
 
 Now suppose that $\Vb_t$ respects $P3$-1 to $P3$-3. Then $g(\Vb_t)$ is, by its definition, such that:
 \beqx
 f^A(S^{A,k}_t,\bar{C}^A_t,D^A_t + g(\Vb_t) = f^B(S^{B,l}_t,\bar{C}^B_t,D^B_t - g(\Vb_t))
 \eeqx
 and then $E_t = g(\Vb_t)$ is the optimal commercial flow. From $P3$-1, $P3$-2 and $P3$-3 we conclude that $\Vb_t \in \Ac_{3,k,l}$ and the strict inequalities in assumptions $P3-2$ and $P3-3$ ensures that $D^A_t + E_t$ and $D^B_t - E_t$ are not discontinuity points of $P^A$ and $P^B$ respectively.
\section{Forward prices computation}
\label{appendix_forward}
This appendix is dedicated to detail the computation of forward prices, especially each term of expressions \eqref{forwardAswitch} and \eqref{forwardBswitch}. In the following, for sake of simpler notation, we consider the specific permutation, noted $\pi_1$, where $S^1_T \le S^2_T \le \dots S^N_T$. The key tool to the computation of all the expectations of equations  \eqref{forwardAswitch} and \eqref{forwardBswitch} is Lemma \ref{lemma_laplace} that allows to compute them by calculating a probability measure, under a multivariate Gaussian probability, of a subspace defined by linear inequalities.

\begin{lemma}
	\label{lemma_laplace}
	Let $X\sim \Nc(\mu,\Sigma)$ be a  $n$-Gaussian vector of mean $\mu$ and covariance $\Sigma$. For $\lambda \in \mathbb{R}^n$ and $f$ a Borel measurable function, we have:  
	\[ 
	\mathbb{E}\left[e^{\lambda ^T X}f(X) \right] = e^{\frac{\lambda^T \Sigma \lambda}{2} + \lambda ^T \mu}~\Ebb \left(f(\tilde{X}) \right)
	\]
	\bigskip 
	with $\tilde{X}\sim \Nc(\mu + \Sigma \lambda;\Sigma)$.
\end{lemma}
\subsection{Gaussian law of $\Vb_T | \Vb_t$}
Due to assumptions in section \ref{sec_model_fuel}, The law of $\Vb_T$ conditional to $\Vb_t$ is Gaussian with mean $\mub(t,T)=\left[\mu_1,\mu_2,\dots,\mu_{N+2}\right]^T$ and covariance $\Sigmab(t,T)=\left( \Sigma_{i,j}\right)_{i,j=1,\dots,N+2}$ defined by:
\beqnx
\mu_n & = & \log S^n_t e^{-a_n(T-t)} + m_n(t)\left( 1-e^{-a_n(T-t)}\right), \quad n=1,\dots,N \\
\mu_{N+1} & = & f^A_T + D^A_t e^{-a^A(T-t)} \\
\mu_{N+2} & = & f^B_T + D^B_t e^{-a^B(T-t)} \\ \\
\Sigma_{i,j} & = & \rho_{i,j}\sigma_i \sigma_j \frac{1-e^{-(a_i+a_j)(T-t)}}{a_i+a_j}, \quad i,j=1,\dots,N+2 \\
\eeqnx
with notations $a_{N+1}=a^A$, $a_{N+2}=a^B$, $\sigma_{N+1}=\sigma^A$ and $\sigma_{N+2}=\sigma^B$.
\noindent
\subsection{The case $\Ac_{1,k,l}$}
\label{cas_A1}
We are first interested in the first term of expression \eqref{forwardAswitch}.  Using Lemma \ref{lemma_laplace} we have:
\beqnx
E_{1,k,l}^A & = & 
\Ebb_t^\Qbb \left[ f^A(S^{A,k}_T,\bar{C}^A_T,D^A_T+\bar{E}) \unbb_{\Ac_{1,k,l}}(\Vb_T) \unbb_{\Sc^{\pi_1}}(\Vb_T)\right] \\
& = & \Ebb_t^\Qbb \left[ e^{\lambda^T \Vb_T + \eta} \unbb_{\Ac_{1,k,l}}(\Vb_T) \unbb_{\Sc^{\pi_1}}(\Vb_T)\right] \\
& = & e^{\frac{\lambda^T \Sigmab(T-t) \lambda}{2} + \lambda ^T \mu(T-t)+\eta}~ \Qbb \left( \tilde{\Vb_T} \in \Ac_{1,k,l} \cap \Sc^{\pi_1} | \Vb_t\right)
\eeqnx
with
\beqx
\lambda  =   \begin{pmatrix}
	E_k^A  \\ 
	-\beta^A \\
	0
\end{pmatrix}, \quad 
\eta  =  \alpha^A + \beta^A (\bar{C}^A_T - \bar{E})
\eeqx
where $E_k^A$ is the $N$-dimensional canonical vector with a unit value at the coordinate corresponding to the production cost $S^{A,k}$,\\ and $\tilde{\Vb_T} |\Vb_t  \sim  \Nc \left(\mub(T-t) + \Sigmab(T-t)\lambda~;~\Sigmab(T-t) \right) $.

Using the definition of $\Mc_{k,l}$ in \eqref{Mkl_switch}, we get:
\beqx
\Ac_{1,k,l} \cap \Sc^{\pi_1} = \Ac_1 \cap \Sc^{\pi_1} \cap \left\{\omega \in \Omega: D^A_t + \bar{E} \in I_t^{A,k} ~;~ D^B_t - \bar{E} \in I_t^{B,l} \right\}
\eeqx

Therefore, the computation of $E_{1,k,l}^A$ needs the computation of the multi-dimensional cumulative distribution related to these inequalities:
\begin{enumerate}[{I}-1 :]
	\item $\sum_{i=0}^{k-1}C^{A,i}_T \le  D^A_T + \bar{E}  < \sum_{i=0}^{k}C^{A,i}_T $
	\item $\sum_{i=0}^{l-1}C^{B,i}_T  \le  D^B_T - \bar{E}< \sum_{i=0}^{l}C^{B,i}_T$
	\item $\log S^{A,k}_T + \alpha^A + \beta^A(\bar{C}^A_T - D^A_T - \bar{E})  <  \log S^{B,l}_T + \alpha^B + \beta^B(\bar{C}^B_T - D^B_T + \bar{E}) $
	\item $ S^1_T \le S^2_T  \le  \dots  \le S^N_T$
\end{enumerate}

More precisely, one needs to compute $\Qbb(\ab \le \Mb \tilde{\Vb}_T \le \bb |\Vb_t)$ with:
\beqnx
\ab & = & \begin{pmatrix}
\sum_{i=0}^{k-1}C^{A,i}_T - \bar{E} \\
\sum_{i=0}^{l-1}C^{B,i}_T + \bar{E} \\
-\infty \\
0 \\
\vdots \\
0
\end{pmatrix} ,  \quad
\bb  =   \begin{pmatrix}
	\sum_{i=0}^{k}C^{A,i}_T - \bar{E} \\
	\sum_{i=0}^{l}C^{B,i}_T + \bar{E} \\
	\alpha^B - \alpha^A + \beta^B(\bar{C}^B_T + \bar{E}) - \beta^A(\bar{C}^A_T - \bar{E}) \\
	+\infty  \\
	\vdots \\
	+\infty 
\end{pmatrix}
\\ \\
\Mb^T & = & \begin{pmatrix}
	\mbox{\textbf{0}} & \mbox{\textbf{0}} & E^A_k - E^B_l & \Db \\
	1 & 0 & -\beta^A      & 0 \\
	0 & 1 & \beta^B       & 0
\end{pmatrix}
\eeqnx
In these expression, the first three elements of $\ab$ and $\bb$ and the first three columns of $\Mb^T$ (with the notation \textbf{0} for a vector of dimension $N$ composed of zeros) are related to inequalities $I$-1 to $I$-3. The $N-1$ last elements of $\ab$ and $\bb$ and the last block of $\Mb^T$ are related to inequalities $I$-4, where $\Db$ is the $(N-1) \times (N-1)$ differentiating matrix:
\beqx
\Db  =  \begin{pmatrix}
	-1      & 0          & \dots  & \dots   & 0 \\
	1      & -1         & 0      & \dots   & 0 \\
	0      &   1        & \ddots &  \ddots & \vdots \\
	\vdots &   \ddots   & \ddots &  \ddots & 0 \\
	\vdots &            & \ddots &  \ddots & -1 \\
	0      &    \dots   & \dots  &   0     & 1  \\ 
\end{pmatrix}
\eeqx

The computation of $\Ebb_t^\Qbb \left[ f^B(S^{B,l}_T,\bar{C}^B_T,D^B_T-\bar{E}) \unbb_{\Ac_{1,k,l}}(\Vb_T) \unbb_{\Sc^{\pi_1}}(\Vb_T)\right]$ is done using the same tools, replacing $\lambda$ and $\eta$ by:
\beqx
\lambda' =   \begin{pmatrix}
	E_l^B  \\ 
	0 \\
	-\beta^B
\end{pmatrix},\quad
\eta'  =  \alpha^B + \beta^B(\bar{C}^B_T + \bar{E})
\eeqx
where $E_l^B$ is the $N$-dimensional canonical vector with a unit value at the coordinate corresponding to the production cost $S^{B,l}$,\\ and $\tilde{\Vb_T} |\Vb_t  \sim  \Nc \left(\mub(T-t) + \Sigmab(T-t)\lambda'~;~\Sigmab(T-t) \right) $, the latter implying that the same event (described by $\ab$, $\bb$ and $\Cb$) must be measured, but under a different Gaussian law.

\subsection{The case $\Ac_{2,k,l}$}
\label{cas_A2}
By using the same arguments of the previous section, the computation of terms $\Ebb_t^\Qbb\left[f^A(S^{A,k}_T,\bar{C}^A_T,D^A_T+\ubar{E}) \unbb_{\Ac_{2,k,l}}(\Vb_T) \unbb_{\Sc^{\pi_1}}(\Vb_T) 
	\right]$
and \\
$\Ebb_t^\Qbb\left[f^B(S^{B,l}_T,\bar{C}^B_T,D^B_T-\ubar{E}) \unbb_{\Ac_{2,k,l}}(\Vb_T) \unbb_{\Sc^{\pi_1}}(\Vb_T) 
\right]$
can be done by the change of probability related to the same $\lambda$ and $\lambda'$ of section \ref{cas_A1}, and:
\beqx
\eta  =  \alpha^A + \beta^A (\bar{C}^A_T - \ubar{E}), \quad
\eta'  =  \alpha^B + \beta^B(\bar{C}^B_T + \ubar{E}),
\eeqx
and the computation, under these probabilities, of the event $\ab \le \Mb \tilde{\Vb}_T \le \bb$ with the same matrix $\Mb$ as in the section B.2 and:
\beqx
\ab  =  \begin{pmatrix}
	\sum_{i=0}^{k-1}C^{A,i}_T - \ubar{E} \\
	\sum_{i=0}^{l-1}C^{B,i}_T + \ubar{E} \\
	\alpha^B - \alpha^A + \beta^B(\bar{C}^B_T + \ubar{E}) - \beta^A(\bar{C}^A_T - \ubar{E}) \\
	0 \\
	\vdots \\
	0
\end{pmatrix} ,  \quad
\bb  =   \begin{pmatrix}
	\sum_{i=0}^{k}C^{A,i}_T - \ubar{E} \\
	\sum_{i=0}^{l}C^{B,i}_T + \ubar{E} \\
	+\infty \\
	+\infty  \\
	\vdots \\
	+\infty 
\end{pmatrix}
\eeqx
\subsection{The case $\Ac_{3,k,l}$}
\label{cas_A3}
In this section the objective is to detail the computation in the case where spot prices have converged in the two markets. We must then decompose the event $\Ac_{3,k,l}$ in 3 events as described in section \ref{sec_spot_price}.
\subsubsection{The case $\Ac_{3,k,l}^A$}
\label{cas_A3A}
In this case, the change of probability is defined with:
\beqx
\lambda  =  \begin{pmatrix} 
	E^B_l\\
	-\beta^B \\
	-\beta^B
\end{pmatrix},\quad
\eta  =  \alpha^B + \beta^B\left( \bar{C}^B_T + \sum_{i=0}^{k-1}C^{A,i}_T \right)
\eeqx
and the event $\ab\le \Mb\tilde{\Vb_T} \le \bb$ defined by:
\beqnx
\ab & = & \begin{pmatrix}
	\sum_{i=0}^{k-1}C^{A,i}_T - \bar{E} \\
	\sum_{i=0}^{k-1}C^{A,i}_T+ \sum_{i=0}^{l-1}C^{B,i}_T  \\
	-\infty \\
	\alpha^B - \alpha^A + \beta^B(\bar{C}^B_T +\sum_{i=0}^{k-1}C^{A,i}_T) - \beta^A(\bar{C}^A_T - \sum_{i=0}^{k-1}C^{A,i}_T) \\
	0 \\
	\vdots \\
	0
\end{pmatrix} \\ \\
\bb  & = &   \begin{pmatrix}
	\sum_{i=0}^{k-1}C^{A,i}_T - \ubar{E} \\
	\sum_{i=0}^{k-1}C^{A,i}_T+ \sum_{i=0}^{l}C^{B,i}_T  \\
	\alpha^B - \alpha^A + \beta^B(\bar{C}^B_T +\sum_{i=0}^{k-1}C^{A,i}_T) - \beta^A(\bar{C}^A_T - \sum_{i=0}^{k-1}C^{A,i}_T) \\
	+\infty \\
	\vdots \\
	+\infty 
\end{pmatrix}
\\ \\
\Mb^T & = & \begin{pmatrix}
	\mbox{\textbf{0}} & \mbox{\textbf{0}} & E^A_{k-1} - E^B_l & E^A_{k} - E^B_l &\Db \\
	1 & 1 & \beta^B &\beta^B     & 0 \\
	0 & 1 & \beta^B &  \beta^B      & 0
\end{pmatrix}
\eeqnx

\subsubsection{The case $\Ac_{3,k,l}^B$}
\label{cas_A3B}
The case $\Ac_{3,k,l}^B$ is the same as $\Ac_{3,k,l}^A$ where one needs to invert market $A$ and market $B$. It gives:
\beqx
\lambda  =  \begin{pmatrix} 
	E^A_k\\
	-\beta^A \\
	-\beta^A
\end{pmatrix},\quad
\eta  =  \alpha^A + \beta^A\left( \bar{C}^A_T + \sum_{i=0}^{l-1}C^{B,i}_T \right)
\eeqx
and the event $\ab\le \Mb\tilde{\Vb_T} \le \bb$ defined by:
\beqnx
\ab & = & \begin{pmatrix}
	\sum_{i=0}^{l-1}C^{B,i}_T + \ubar{E} \\
	 \sum_{i=0}^{l-1}C^{B,i}_T+ \sum_{i=0}^{k-1}C^{A,i}_T  \\
	-\infty \\
\alpha^A - \alpha^B + \beta^A(\bar{C}^A_T +\sum_{i=0}^{l-1}C^{B,i}_T) - \beta^B(\bar{C}^B_T - \sum_{i=0}^{l-1}C^{B,i}_T) \\
	0 \\
	\vdots \\
	0
\end{pmatrix} \\ \\
\bb  & = &   \begin{pmatrix}
	\sum_{i=0}^{l-1}C^{B,i}_T + \bar{E} \\
	 \sum_{i=0}^{l-1}C^{B,i}_T+ \sum_{i=0}^{k}C^{A,i}_T  \\
	\alpha^A - \alpha^B + \beta^A(\bar{C}^A_T +\sum_{i=0}^{l-1}C^{B,i}_T) - \beta^B(\bar{C}^B_T - \sum_{i=0}^{l-1}C^{B,i}_T) \\
	+\infty \\
	\vdots \\
	+\infty 
\end{pmatrix}
\\ \\
\Mb^T & = & \begin{pmatrix}
	\mbox{\textbf{0}} & \mbox{\textbf{0}} & E^B_{l-1} - E^A_k & E^B_{l} - E^A_k &\Db \\
	0 & 1 & \beta^A &\beta^A     & 0 \\
	1 & 1 & \beta^A &  \beta^A      & 0
\end{pmatrix}
\eeqnx

\subsubsection{The case $\Ac_{3,k,l}^C$}
\label{cas_A3C}
In this case, the change of probability is defined with:
\beqnx
\lambda & = & \begin{pmatrix} 
	\frac{\beta^B}{\beta^A + \beta^B}E^A_k +\frac{\beta^A}{\beta^A + \beta^B} E^B_l) \\
	-\frac{\beta^B}{\beta^A + \beta^B} \\
	-\frac{\beta^A}{\beta^A + \beta^B}%
	\end{pmatrix}\\
\eta & = & \frac{1}{\beta^A + \beta^B}\left(\beta^B\alpha^A + \beta^A \alpha^B +\beta^A \beta^B (\bar{C}^A_T + \bar{C}^B_T ) \right)
\eeqnx
and the event $\ab\le \Mb\tilde{\Vb_T} \le \bb$ defined by:

\beqnx
\ab & = & \begin{pmatrix}
	(\beta^A + \beta^B)\ubar{E} - K \\
	
	(\beta^A + \beta^B)\sum_{i=0}^{k-1}C^{A,i}_T- K  \\
	
	(\beta^A + \beta^B)\sum_{i=0}^{l-1}C^{B,i}_T+ K \\
	
	0 \\
	
	\vdots \\
	
	0
\end{pmatrix} \\ \\
\bb  & = &   \begin{pmatrix}
	(\beta^A + \beta^B)\bar{E} - K \\
	
	(\beta^A + \beta^B)\sum_{i=0}^{k}C^{A,i}_T- K  \\
	
	(\beta^A + \beta^B)\sum_{i=0}^{l}C^{B,i}_T+ K \\
	
	+\infty \\
	\vdots \\
	+\infty 
\end{pmatrix}
\\ \\
\Mb^T & = & \begin{pmatrix}
	E^A_k-E^B_l & E^A_k-E^B_l & E^B_l-E^A_k  &\Db \\
	
	-\beta^A & \beta^B & \beta^A     & 0 \\
	
	\beta^B & \beta^B & \beta^A      & 0
\end{pmatrix}
\eeqnx
with $K=\alpha^A-\alpha^B+\beta^A \bar{C}^A_T - \beta^B \bar{C}^B_T$.

\end{document}